\documentclass[prb,twocolumn,showpacs]{revtex4}
\usepackage{epsfig}
\usepackage{dcolumn}
\usepackage{epstopdf}
\usepackage{graphicx}
\usepackage{subfigure}
\usepackage{epsf}
\begin{document}
\title{A variational pseudo-self-interaction correction approach: ab-initio description of correlated 
oxides and molecules}
\author{A. Filippetti$^1$}
\author{C. D. Pemmaraju$^2$}
\author{P. Delugas$^1$}
\author{D. Puggioni$^1$}
\author{V. Fiorentini$^1$}
\author{S. Sanvito$^2$}
\affiliation{1) CNR-IOM UOS Cagliari, and Dipartimento di Fisica, Universit\`a di Cagliari, I-09042 Monserrato (CA), Italy}
\affiliation{2) School of Physics and CRANN, Trinity College Dublin 2, Ireland}

\date{\today}
\begin{abstract}
We present a fully variational generalization of the pseudo self-interaction correction (VPSIC) approach previously 
presented in two implementations based on plane-waves and atomic orbital basis set, known as PSIC and ASIC, respectively. 
The new method is essentially equivalent to the previous version for what concern the electronic properties, but it can be
exploited to calculate total-energy derived properties as well, such as forces and structural optimization. We apply the 
method to a variety of test cases including both non-magnetic and magnetic correlated oxides and molecules, showing a 
generally good accuracy in the description of both structural and electronic properties. 
\end{abstract}

\maketitle

\section{Introduction}

The ab-initio, density-functional theory (DFT) based determination of structural and electronic properties of correlated
systems remains up today an outstanding challenge. The most relevant bottleneck rests in the fact that advanced approaches
that are general and powerful enough to tackle the description of strong-correlated systems are also usually heavily 
demanding in terms of required computing resources. This limits the system size that can be practically afforded 
to few tents of atoms per unit cell. On the other hand, most of the interesting behaviours encountered in correlated 
systems requires large cell size already at the level of bulk properties, due the possible coexistence of several 
juxtapposed orderings (structural, magnetic, orbital and charge ordering). Furthermore, several much celebrated 
phenomenologies involve doping (high-Tc superconductivity in cuprates, colossal magnetoresistivity in manganites, 
magnetic ordering in diluted semiconductors, etc.) whose treatment at generic concentration is a formidable computing 
task; finally, oxide interfaces and multilayers which are ground of recent intriguing discoveries such as 2D electron gas 
behaviour\cite{ohtomo} may require even larger simulation effort. 

As a matter of fact, we are in the need of treating about one or two hundred atom systems, a size that can be only 
afforded at a computational cost substantially similar to that of standard local (spin)density functional (L(S)DA)
theory or its generalized gradient (GGA) version. Beyond-LSDA approaches which are agile enough to satisfy this requirement
are very few: the very popular LDA+U\cite{anisimov} is certainly one of those; a later approach with similar characteristics 
is the PSIC, implemented in the past in two different setting: plane-wave and ultrasoft pseudopotential basis 
set\cite{fs,ff_coll} in the framework of the home-made PWSIC code; local orbital and pseudopotential basis set 
(also called ASIC)\cite{pemmaraju} in the framework of the SIESTA code. LDA+U and PSIC/ASIC move from different 
conceptual viewpoints: the former introduces an effective Coulomb repulsion which is tipically mistreated in LSDA; 
the latter subtracts from the LSDA functional an approximate (i.e. atomic orbital-averaged) self-interaction (SI), that 
is the unphysical interaction of an electron with its own generated potential.

Despite this difference, the two theories act in similar way, i.e. correcting the LSDA eigenvalues by a quantity linearly 
dependent on the orbital occupations; in fact, the PSIC can be substantially viewed as an all-orbital generalization of 
the LDA+U, but with two relevant advantages over the latter: the capability of curing the LSDA failures in more general 
situations (i.e. not limited to magnetic and insulating systems) and the absence of an explicit parametric dependence.
In PSIC/ASIC, indeed, the function of U si replaced by the atomic self-interaction potentials, which are extracted 
from the free atom (thus they are universal, i.e. only dependent on the atomic species of the systems) and then plug into
the band structure Hamiltonian. At variance with more fundamental SI removal strategies such as the original 
Perdew-Zunger approach\cite{pz} (PZ-SIC), or the later generalization to extended systems\cite{svane,szotek} which 
draw on dramatic complications in formalism and conceptual interpretations\cite{stengel,ff_coll}, the PSIC keep all the 
simplicity tipical of LDA/GGA: a single-particle potential which is orbital-independent and translationally invariant 
(i.e. obeying Bloch symmetry), and energy functional invariant under unitary rotation of the occupied eigenstates.

The PSIC/ASIC approach demonstrated a consistent accuracy in the description of the electronic properties for a vast range
of systems\cite{ff_coll}, at a computational cost not much larger than that of the LSDA itself (in particular the ASIC
is the implementation of choice for large-size systems as it can easily afford few hundreds atom simulations, while the 
heavier PSIC can be seen as the standard reference for what concern the methodological accuracy). Despite these virtues, 
PSIC and ASIC have had a relatively small following with respect to the LDA+U so far, mainly in reason of the serious 
hindrance represented by the lack of variationality: in Ref.\onlinecite{fs}the PSIC potential is in fact generated as an 
ansatz on the single-particle Kohn-Sham (KS) potential, not from a germinal energy functional. This shortcoming precludes,
in practice, the access to ground-state total energy properties.    

In this paper we point to overcome this limitation, presenting a fully variational version of PSIC/ASIC (hereafter
called VPSIC, to be distinguished by their previous siblings) built out of an energy functional which recovers, by
Euler-Lagrange derivative, a set of single-particle KS equations essentially similar to those of the elder PSIC/ASIC 
scheme, thus keeping the successful description of the electronic properties, but adding up the possibility to deliver 
all those ground-state properties expected by standard ab-initio theories. 

Here we describe the new formulations and give wide evidence of the aforementioned capabilities presenting results for a 
range of bulks (including non magnetic and magnetic titanates and manganites) and molecules. Extended systems are treated
using the plane-wave basis implementation, the molecules by the atomic orbital basis implementation. We remark that other 
VPSIC results have been already presented in separate recent publications: a lenghtly description of the properties of 
transition metal monoxides\cite{tmo}; a detailed account of the properties of 2D electron gas formation at the 
SrTiO$_3$/LaAlO$_3$ interface\cite{delugas}; together with the present article, they furnish a solid evidence of the 
effectiveness of the new method in the study of generic systems characterized by strong to moderate electron correlation.  

The article is organized as it follows: in Section \ref{form} the general formulation is illustrated; in 
Section \ref{res_ext} results for non magnetic oxide insulators (Subsection \ref{res_para}), titanates (Subsection 
\ref{res_tita}) and manganites (Subsection \ref{res_manga}) are presented. Section \ref{res_mol} is devoted to illustrate
results for molecules. Finally, in Section \ref{concl} we draw our summary and conclusions. Several specific implementations
are included in the Appendices: in \ref{app_1} and \ref{app_2} we present an extension of the VPSIC energy functional and 
forces formulations, respectively, for the case of ultrasoft pseudopotentials (USPP), while \ref{app_3} and \ref{app_4}
are dedicated to the description of VPSIC energy functional and forces formulation specific for atomic orbital 
basis set.

\section{Variational pseudo-self-interaction formulation}
\label{form}

In this Section we present the generic variational formulation, not related to any specific
basis function implementation.

\subsection{VPSIC energy functional and related Kohn-Sham equations}
\label{form_general}

We start from the following VPSIC energy functional:
 
\begin{equation}
E^{VPSIC}[\{\psi\}]\,=\,E^{LSD}[\{\psi\}]-{1\over 2}\sum_{ij\nu\sigma}\,{\cal E}^{SI}_{ij\sigma\nu}\,P^{\sigma}_{ji\nu}
\label{vpsic_etot}
\end{equation}

where $E^{LSD}$ is the usual LSDA energy functional:

\begin{eqnarray}
E^{LSD}[\{\psi\}] = T_s[\{\psi\}]+E_H[n]+E_{xc}[n_+,n_-]+E_{ion}[\{\psi\}]
\nonumber
\end{eqnarray}

written as sum of (non-interacting) kinetic (T$_s$), Hartree (E$_H$), exchange-correlation (E$_{xc}$), and electron-ion 
(E$_{ion}$) energies ($\psi$ are single-particle wavefunctions, n$_+$ and n$_-$ up- and down-polarized electron densities, 
and n=n$_+$+n$_-$). Eq.\ref{vpsic_etot} follows the spirit of the original Perdew-Zunger procedure\cite{pz} (hereafter called PZ-SIC),
and subtracts from the LSDA total energy a SI term written as a sum of effective single-particle SI energies (${\cal E}^{SI}$) rescaled 
by some orbital occupations $P$. Here $i$, $j$ are sets ($l_i$,$m_i$, $l_i$,$m_j$) of atomic quantum numbers (typically relative to
a minimal atomic wavefunction basis set) while $\sigma$ and $\nu$ indicate spin and atomic site, respectively (non-diagonal formulation is 
necessary to enforce covariancy, thus $i$=($l_i$,$m_i$), $j$=($l_i$,$m_j$)). 

Most of the peculiarity of the VPSIC approach resides in the way the second term of Eq.\ref{vpsic_etot} is written for an
extended system whose eigenfunctions are Bloch states ($\psi_{n{\bf k}}^{\sigma}$). The orbital occupations are then 
calculated as projection of Bloch states onto localized (atomic) orbitals (hereafter indicated as $\{\phi\}$):

\begin{eqnarray}
P^{\sigma}_{ij\nu}=\,\sum_{n{\bf k}}\, f_{n{\bf k}}^{\sigma}\, \langle\psi_{n{\bf k}}^{\sigma}|
\phi_{i,\nu}\rangle \> \langle \phi_{j,\nu} |\psi_{n{\bf k}}^{\sigma}\rangle,
\label{occ}
\end{eqnarray}

where $f_{n{\bf k}}^{\sigma}$ are Fermi occupancies. For the effective SI energies we adopt a similar approach: 

\begin{eqnarray}
{\cal E}^{SI}_{ij\sigma\nu}\,=\,\sum_{n{\bf k}}\, f_{n{\bf k}}^{\sigma}\, \langle\psi_{n{\bf k}}^{\sigma}|
\gamma_{i,\nu}\rangle C_{ij} \langle\,\gamma_{j,\nu}|\psi_{n{\bf k}}^{\sigma}\rangle
\label{epsi-si}
\end{eqnarray}

where $\gamma_{i,\nu}$ is the projection function associated to the SI potential of the $i^{th}$ atomic orbital centered
on atom $\nu$:
 
\begin{eqnarray}
\gamma_{i\nu}({\bf r}-{\bf R}_{\nu})\>=\> V_{Hxc}[n_{i\nu}({\bf r}-{\bf R}_{\nu}),0]\,\phi_{i\nu}({\bf r}-{\bf R}_{\nu}),
\label{gamma}
\end{eqnarray}

where $n_{i\nu}({\bf r})$=$\phi^2_{i\nu}({\bf r})$. We express the Hartree plus exchange-correlation atomic SI potential $V_{Hxc}$ 
in radial approximation: $V_{Hxc}$ = $V_H[n_{l_i\nu}]+V_{xc}[n_{l_i\nu},0]$ = $\partial E_{Hxc}[n_{l_i\nu}]/\partial n_{l_i\nu}$ 
(calculated at full polarization: n=n$_+$, n$_-$=0). Finally, the $C_{ij}$ are normalization coefficients:

\begin{eqnarray}
C^{-1}_{l_i,m_i,m_j}=\int d{\bf r}\>\phi_{l_i m_i}({\bf r})\,V_{Hxc}[n_{l_i,\nu}(r),0]\,\phi_{l_i m_j}({\bf r})
\label{cij}
\end{eqnarray}

with $l_i$=$l_j$. They are purely atomic and do not depend on the atomic positions. The use of projectors $\gamma$ in 
Eq.\ref{epsi-si} is aimed at casting the SI energy in fully non-local form (analogous to the fully-non local 
pseudopotential form due to Kleinman and Bylander\cite{kb}) which consent a huge saving of computational effort 
when calculated in reciprocal space.

To grab the idea behind Equations \ref{epsi-si}, \ref{gamma}, and \ref{cij}, notice that in the limit of large 
atomic separation, the Bloch states $\psi_{n{\bf k}}$ are brought back to atomic orbitals $\phi_{i\nu}$, and 
${\cal E}^{SI}_{ij\sigma\nu}$ to atomic SI energies $\epsilon^{SI}_i$ (discarding spin and atomic index):

\begin{eqnarray}
\epsilon^{SI}_i\,=\,\int \, d{\bf r}\> n_i({\bf r})\,\left( V_H[n_i({\bf r})]+ V_{xc}[n_i({\bf r}),0] \right)
\label{e_at}
\end{eqnarray}

Thus, the orbital occupations $P^{\sigma}_{ij\nu}$ (if suitably normalized) act as scaling factors for the atomic SI 
energies, assumed as the upper limit of the SI correction amplitude. 

We remark that in the atomic limit Eq.\ref{vpsic_etot} goes back to the PZ-SIC total energy espression only for what 
concern the Hartree SI part, while our SI exchange-correlation energy density ((1/2)$V_{xc}[n_i,0]$) departs from the 
PZ-SIC espression ${\epsilon}_{xc}[n_i,0]$, since $V_{xc}={\epsilon}_{xc}+n_i\partial{\epsilon}_{xc}/\partial n_i$). 

From Eq.\ref{vpsic_etot} we obtain the corresponding VPSIC KS Equations through the usual Euler-Lagrange 
derivative:

\begin{eqnarray}
{\partial E^{VPSIC}\over \partial \psi^*_{nk\sigma}}\,=\,\tilde{\epsilon}_{nk\sigma}\,\psi_{nk\sigma}
\nonumber
\end{eqnarray}
\begin{eqnarray}
=\,\hat{h}^{LSD}_{\sigma}\psi_{nk\sigma}
-{1\over 2}\sum_{ij\nu}\,\left\{
{\partial {\cal E}^{SI}_{ij\sigma\nu}\over \partial \psi^*_{nk\sigma}}  P^{\sigma}_{ji\nu}
+ {\cal E}^{SI}_{ij\sigma\nu}  {\partial P^{\sigma}_{ji\nu} \over \partial \psi^*_{nk\sigma}} 
\right\}
\label{ks-sic}
\end{eqnarray}
  
where $\tilde{\epsilon}_{nk\sigma}$ are VPSIC eigenvalues, and;

\begin{eqnarray}
h^{LSD}_{\sigma}({\bf r})\,=\,-{ \nabla^2_{{\bf r}} \over 2} 
\nonumber
\end{eqnarray}
\begin{eqnarray}
+V_H[n({\bf r})]+V_{xc}[n_+({\bf r}),n_-({\bf r})]+V_{ion}({\bf r})
\label{h_lsd}
\end{eqnarray}

is the usual KS LSDA Hamiltonian, and: 

\begin{eqnarray}
{\partial {\cal E}^{SI}_{ij\sigma\nu}\over \partial \psi^*_{nk\sigma}} \,=\,
|\gamma_{i,\nu}\rangle C_{ij\nu} \langle\,\gamma_{j,\nu}|\psi_{n{\bf k}}^{\sigma}\rangle;
\label{depsi}
\end{eqnarray}

\begin{eqnarray}
{\partial P^{\sigma}_{ji\nu} \over \partial \psi^*_{nk\sigma}} \,=\,
|\phi_{j,\nu}\rangle \> \langle \phi_{i,\nu}|\psi_{n{\bf k}}^{\sigma}\rangle.
\label{docc}
\end{eqnarray}

The first sum term in curl parenthesis in Eq.\ref{ks-sic} corresponds to the SI potential projector written as in
the original VPSIC KS Equations. Since the two sums in curl parhenthesis are two ways to describe substantially the same
physical quantity (i.e. the SI potential), in practice they give similar results when applied onto the Bloch
state. It follows that Eq.\ref{ks-sic} describes an energy spectrum substantially similar to that of the non variational
scheme, but with the added bonus to derive from the VPSIC energy functional.
   
In DFT methods it is customary to rewrite the total energy in terms of eigenvalue sum. 
Indicating with $\epsilon_{nk\sigma}$ the LSDA eigenvalues, it is immediate to verify that:

\begin{eqnarray}
\sum_{n{\bf k}\sigma}\, f_{n{\bf k}}^{\sigma}\,\tilde{\epsilon}_{nk\sigma}\,=\,
\sum_{n{\bf k}\sigma}\, f_{n{\bf k}}^{\sigma}\, \langle\psi_{n{\bf k}}^{\sigma}|
{\partial E^{VPSIC}\over \partial \psi^*_{nk\sigma}}\rangle\,
\nonumber
\end{eqnarray}
\begin{eqnarray}
=\,\sum_{n{\bf k}\sigma}\, f_{n{\bf k}}^{\sigma}\, \epsilon_{nk\sigma}
-\sum_{ij\sigma\nu}\,{\cal E}^{SI}_{ij\sigma\nu} P^{\sigma}_{ji\nu},
\label{eig-sum}
\end{eqnarray}

then Eq. \ref{vpsic_etot} can be rewritten as:

\begin{equation}
E^{VPSIC}[\{\psi\}]\,=\,\tilde{E}^{LSD}[\{\psi\}]+{1\over 2}\sum_{ij\sigma\nu}\,
{\cal E}^{SI}_{ij\sigma\nu}\,P^{\sigma}_{ji\nu}
\label{esic2}
\end{equation}

where: 

\begin{eqnarray}
\tilde{E}^{LSD}[\{\psi\}] = \sum_{n{\bf k}\sigma}\, f_{n{\bf k}}^{\sigma}\, \tilde{\epsilon}_{nk\sigma}
+E_{Hxc}[n_+({\bf r}),n_-({\bf r})]
\nonumber
\end{eqnarray}
\begin{eqnarray}
+E_{ion}-\sum_{\sigma}\int d{\bf r} n_{\sigma}({\bf r})V^{\sigma}_{Hxc}[n_+({\bf r}),n_-({\bf r})]
\label{esic3}
\end{eqnarray}

is the LSDA energy functional but now including the VPSIC eigenvalues in place of the LSDA eigenvalues.
Finally, in the original PSIC formulation the SI $V_{Hxc}$ potential is rescaled by a relaxation factor $\alpha$=1/2, 
to take into account the screening (i.e. the suppression) of atomic self-interaction caused by the surrounding charge 
of the other atoms (see Ref.\cite{ff_coll} for an extended discussion). A careful testing on a large series of compounds
\cite{pemmaraju} confirmed that this relaxation value is the most adequate for almost all the examined bulk systems, 
whereas for molecules the atomic SI ($\alpha$=1) is the most appropriate choice. We mantain this recipe in the present 
formulation, using the 1/2 factor only for calculations related to extended systems.  

%%%%%%%%%%%%%%%%%%%%%%%%%%%%%%%%%%%%%%%%%%%%%%%%%%%%%%%%%%%%%%%%%%%%%%%%%%%%%%%%%%%%%%%%%%%%%%
%%%%%%%%%%%%%%%%%%%%%%%%%%%%%%%%%%%%%%%%%%%%%%%%%%%%%%%%%%%%%%%%%%%%%%%%%%%%%%%%%%%%%%%%%%%%%%

\subsection{Simplified variants of VPSIC and relation with the original non-variational method}

From the general espression of Eq.\ref{vpsic_etot} two interesting subcases can be obtained: assumig fixed (i.e. non 
self-consistent) orbital occupations $P_{ij}$, in Eq.\ref{ks-sic} the second term in curl brackets vanishes and the 
VPSIC-KS equations reduce to those of the original PSIC scheme of Ref.\onlinecite{fs} (indeed, it was previously pointed 
out\cite{ff_coll} that the original scheme becomes variational at fixed orbital occupations). 

Another useful subcase is obtained replacing Eq.\ref{epsi-si} with a simplified expression:

\begin{equation}
{\cal E}^{SI}_{ij\sigma\nu}\,=\,P^{\sigma}_{ij\nu} \epsilon^{SI}_{i\nu}\,=\,P^{\sigma}_{l_i m_i m_j\nu} 
\,\epsilon^{SI}_{l_i\nu}
\label{esic-simple}
\end{equation}

where the atomic $\epsilon^{SI}_{l_i\nu}$ (in radial approximation) is given by Eq.\ref{e_at}. Using Eq.\ref{esic-simple}, 
previous Eqs.\ref{vpsic_etot} and \ref{ks-sic}:

\begin{equation}
E^{VPSIC_0}[\{\psi\}]\,=\,E^{LSD}[\{\psi\}]-{1\over 2}\sum_{ij\nu\sigma}\, 
P^{\sigma}_{ij\nu}\,P^{\sigma}_{ji\nu}\,\epsilon^{SI}_{j\nu} 
\nonumber
\end{equation}
\begin{equation}
=\tilde{E}^{LSD}[\{\psi\}]+{1\over 2}\sum_{ij\nu\sigma}\, P^{\sigma}_{ij\nu}\,P^{\sigma}_{ji\nu}\,\epsilon^{SI}_{j\nu} 
\label{avpsic_etot}
\end{equation}

\begin{equation}
{\partial E^{VPSIC_0}\over \partial \psi^*_{nk\sigma}}\,=\,\hat{h}^{LSD}_{\sigma}\psi_{nk\sigma}
-\sum_{ij\nu}\, P^{\sigma}_{ij\nu}{\partial P^{\sigma}_{ji\nu} \over \partial \psi^*_{nk\sigma}}
\epsilon^{SI}_{j\nu} 
\label{ks-vpsic0}
\end{equation}

These simplified VPSIC formalism (hereafter indicated as VPSIC$_0$) is a computationally convenient alternative (especially
in terms of required memory) to perform structural optimizations in large size-systems. In the results Section we will give
evidence that at least in case of non-magnetic semiconductors and insulators it furnishes electronic and structural 
properties in good accord with the VPSIC. However it is tipically less satisfying for the electronic properties of 
magnetic systems.

%%%%%%%%%%%%%%%%%%%%%%%%%%%%%%%%%%%%%%%%%%%%%%%%%%%%%%%%%%%%%%%%%%%%%%%%%%%%%%%%%%%%%%%%%
%%%%%%%%%%%%%%%%%%%%%%%%%%%%%%%%%%%%%%%%%%%%%%%%%%%%%%%%%%%%%%%%%%%%%%%%%%%%%%%%%%%%%%%%%%
\subsection{Forces formulation}
%%%%%%%%%%%%%%%%%%%%%%%%%%%%%%%%%%%%%%%%%%%%%%%%%%%%%%%%%%%%%%%%%%%%%%%%%%%%%%%%%%%%%%%%%
%%%%%%%%%%%%%%%%%%%%%%%%%%%%%%%%%%%%%%%%%%%%%%%%%%%%%%%%%%%%%%%%%%%%%%%%%%%%%%%%%%%%%%%%%

In VPSIC the atomic forces formulation follows from the usual Hellmann-Feynmann procedure. It is obtained as the 
LSDA expression augmented by a further additive contribution due to the atomic-site dependence of the SI projectors:

\begin{eqnarray}
-{\partial E^{VPSIC}[\{\psi\}]\over \partial {\bf R}_{\nu}}\,=\,{\bf F}_{\nu}^{LSD}+
\nonumber
\end{eqnarray}
\begin{eqnarray}
+{1\over 2}\sum_{ij,nk\sigma}\,f_{n{\bf k}}^{\sigma}\, \left\{ \langle\psi_{n{\bf k}}^{\sigma}| 
{ \partial \gamma_{i,\nu}\over \partial {\bf R}_{\nu} } \rangle 
C_{ij} \langle\,\gamma_{j,\nu}|\psi_{n{\bf k}}^{\sigma}\rangle P^{\sigma}_{ji\nu}[\{\psi\}]+ c.c. \right\}
\nonumber
\end{eqnarray}
\begin{eqnarray}
+{1\over 2}\sum_{ij,nk\sigma}\,f_{n{\bf k}}^{\sigma}\, \left\{ {\cal E}^{SI}_{ij\sigma\nu}[\{\psi\}]\,
\langle\psi_{n{\bf k}}^{\sigma}| { \partial \phi_{i,\nu}\over \partial {\bf R}_{\nu} } \rangle 
\langle\,\phi_{j,\nu}|\psi_{n{\bf k}}^{\sigma}\rangle +c.c \right\}
\label{forces}
\end{eqnarray}

whereas in the simplified version, we have, in addition to ${\bf F}_{\nu}^{LSD}$ the quantity:

\begin{eqnarray}
\sum_{ij,nk\sigma}\,f_{n{\bf k}}^{\sigma}\, \left\{ P^{\sigma}_{ij\nu} \epsilon^{SI}_{j\nu}\,
\langle\psi_{n{\bf k}}^{\sigma}| { \partial \phi_{j,\nu}\over \partial {\bf R}_{\nu} } \rangle 
\langle\,\phi_{i,\nu}|\psi_{n{\bf k}}^{\sigma}\rangle +c.c \right\}
\label{asi_forces}
\end{eqnarray}
 
In writing Eqs.\ref{forces} and \ref{asi_forces}, we have assumed that the force on a given atom $\nu$ only depends on 
the single atomic projector centered on $\nu$. This is not necessarely true if the orbital occupations are to be 
re-orthonormalized on the cell. This choice, which complicates sensitively the above formulation, will be discussed 
in detail in the Appendices, together with the generalization of the method to either plane-waves plus USPP 
or local-orbital plus pseudopotential approach.

%%%%%%%%%%%%%%%%%%%%%%%%%%%%%%%%%%%%%%%%%%%%%%%   RISULTATI  %%%%%%%%%%%%%%%%%%%%%%%%%%%%%%%%%%%%%
%%%%%%%%%%%%%%%%%%%%%%%%%%%%%%%%%%%%%%%%%%%%%%%%%%%%%%%%%%%%%%%%%%%%%%%%%%%%%%%%%%%%%%%%%%%%%%%%%%%

\section{Results: extended systems}
\label{res_ext}

%%%%%%%%%%%%%%%%%%%%%%%%%%%%%%%%%%%%%%%%%%%%%%%%%%%%%%%%%%%%%%%%%%%%%%%%%%%%%%%%%%%%%%%%%%%%%%%%%%%
%%%%%%%%%%%%%%%%%%%%%%%%%%%%%%%%%%%%%%%%%%%%%%%%%%%%%%%%%%%%%%%%%%%%%%%%%%%%%%%%%%%%%%%%%%%%%%%%%%%

We have pointed out in Section \ref{form} that there is a substantial formal similarity between the KS Equations 
derived for the VPSIC and the elder PSIC/ASIC KS equations. Our tests assess that the long series of 
results for the electronic properties obtained with the latter in the last few years remain valid even in 
the framework of the new theory, which gives small differences in e.g. band energies and density 
of states. Hereafter we consider as test cases for VPSIC, materials either never tackled before (it is 
the case of titanates), or afforded in the past (e.g. CaMnO$_3$) but now revisited to specifically address total
energy-derived properties, such as equilibrium structure and magnetic exchange-interactios.

Specifically, we selected three classes of systems which well represent the broad spectrum of the VPSIC capability, 
and at the same time are interesting compounds in itself either from conceptual or technological viewpoint: wide-gap 
oxide insulators, magnetic titanates representative of 3d t$_{2g}$ Mott-insulating perovskites, and magnetic manganites 
representative of 3d e$_g$ charge-transfer insulating perovskites.  

Regarding our technicalities, calculations are carried out with ultrasoft pseudopotentials \cite{uspp} and a 
plane-wave basis set with cut off ranging from 30 to 35 Ryd depending on the specific system, 6$\times$6$\times$6 
special k-point grids for self-consistent calculations, 10$\times$10$\times$10 special k-points and linear tetrahedron
interpolation method for density of states. The Ceperley-Alder-Perdew-Zunger local-density approximation is used for 
the exchange-correlation functional. Structural relaxations are carried out with a convergency threshold of 1 mRy/Bohr 
on the calculated forces.

%%%%%%%%%%%%%%%%%%%%%%%%%%%%%%%%%%%%%%%%%%%%%%%%%%%%%%%%%%%%%%%%%%%%%%%%%%%%%%%%%%%%%%%%%%%%%%%%%%%%%%%%%
%%%%%%%%%%%%%%%%%%%%%%%%%%%%%%%%%%%%%%%%%%%%%%%%%%%%%%%%%%%%%%%%%%%%%%%%%%%%%%%%%%%%%%%%%%%%%%%%%%%%%%%%%
%%%%%%%%%%%%%%%%%%%%%%%%%%%%%%%%%%%%%%%%%%%%%%%%%%%%%%%%%%%%%%%%%%%%%%%%%%%%%%%%%%%%%%%%%%%%%%%%%%%%%%%%%%
\subsection{Wide-gap insulators: LaAlO$_3$, SrTiO$_3$, TiO$_2$}
\label{res_para}
%%%%%%%%%%%%%%%%%%%%%%%%%%%%%%%%%%%%%%%%%%%%%%%%%%%%%%%%%%%%%%%%%%%%%%%%%%%%%%%%%%%%%%%%%%%%%%%%%%%%%%%%%%%%
%%%%%%%%%%%%%%%%%%%%%%%%%%%%%%%%%%%%%%%%%%%%%%%%%%%%%%%%%%%%%%%%%%%%%%%%%%%%%%%%%%%%%%%%%%%%%%%%%%%%%%%%%%
%%%%%%%%%%%%%%%%%%%%%%%%%%%%%%%%%%%%%%%%%%%%%%%%%%%%%%%%%%%%%%%%%%%%%%%%%%%%%%%%%%%%%%%%%%%%%%%%%%%%%%%%%%

As prototypes of non-magnetic wide-gap insulators we selected TiO$_2$ rutile and two perovskite oxides, LaAlO$_3$ and 
SrTiO$_3$ in their cubic bulk structure. Nowadays these materials are ravely investigate for their potential use in 
the field of functional oxides, thus the abundance of theoretical and experimental results make them ideal test cases for
innovative theoretical methods. 

For non magnetic oxides the level of accuracy of standard LDA calculations may vary according to the examined property:
typically a good rendition of structural properties is juxtaposed to an unsatisfactory match of the calculated band 
structure and interband transition energies, involving the well-known underestimation of the fundamental band gap, and 
the poor description of transition-metal $d$ and (on a minor extent) oxygen $p$ states. Our results will demonstrate that 
the VPSIC achieves a sistematical improvement of the electronic properties over LSDA, while preserving a substantially 
similar accuracy for what concerns structural properties. However, it must be kept in mind that the detail of the results 
may (and usually does) depend crucially on a number of computational technicalities typical of our supercell simulations, 
which goes beyond the formalism (e.g. the type of wavefunction basis set, the type of used pseudopotentials, etc.). Thus, 
in order to reach an unbiased evaluation it is always useful to refer the PSIC results not only to the experiment values 
but even to their LDA counterpart, obtained in condition of identical technicalities.
 
%%%%%%%%%%%%%%%%%%%%%%%%%%%%%%%%%%%%%%%%%%%%%%%%%%%%%%%%%%%%%%%%%%%%%%%%%%%%%%%%%%%%%%%%%%
\begin{table}
\caption{Direct ($\Delta$E$_d$) and indirect ($\Delta$E$_i$) band gap energies and O 2p manifold valence bandwidth 
(W$_Op$) in eV, calculated within LDA, PSIC, and PSIC$_0$, compared to the experimental values.
\label{tab_gap}}
\centering\begin{tabular}{|cccccc|}
\hline
            &                               & LDA & PSIC & PSIC$_0$ & Expt. \\
\hline\hline
 TiO$_2$   &     $\Delta$E$_i$ ($\Gamma$-L) & 1.88  & 2.90   & 2.59  &        \\ 
           &     $\Delta$E$_d$ ($\Gamma$)   & 1.84  & 2.93   & 2.62  & [3.0,3.1]\cite{crone, moch}    \\
           &       W$_Op$                   &  6.0  & 6.5    & 5.7   & $\sim$ 7\cite{riga}    \\
 \hline
 SrTiO$_3$ &     $\Delta$E$_i$ (M-$\Gamma$) & 1.69  & 2.94  & 2.62  & 3.25\cite{benthem}   \\
           &     $\Delta$E$_d$ ($\Gamma$)   & 2.04  & 3.30  & 2.95  & 3.75\cite{benthem}    \\
           &       W$_Op$                   & 5.0   & 5.5   & 4.8   & $\sim$ 6\cite{yoshimatsu} \\
\hline
 LaAlO$_3$ &     $\Delta$E$_i$ (M-$\Gamma$) & 2.83  & 5.23  & 4.61  &        \\ 
           &     $\Delta$E$_d$  ($\Gamma$)  & 3.17  & 5.51  & 4.89  & 5.6    \\
           &       W$_Op$                   & 7.6   & 8.38  & 7.27  & $\sim$ 8-9\cite{yoshimatsu}    \\
\hline
\end{tabular}
\end{table}
%%%%%%%%%%%%%%%%%%%%%%%%%%%%%%%%%%%%%%%%%%%%%%%%%%%%%%%%%%%%%%%%%%%%%%%%%%%%%%%%%%%

\begin{figure}
\centerline{\includegraphics[clip,width=7.0cm]{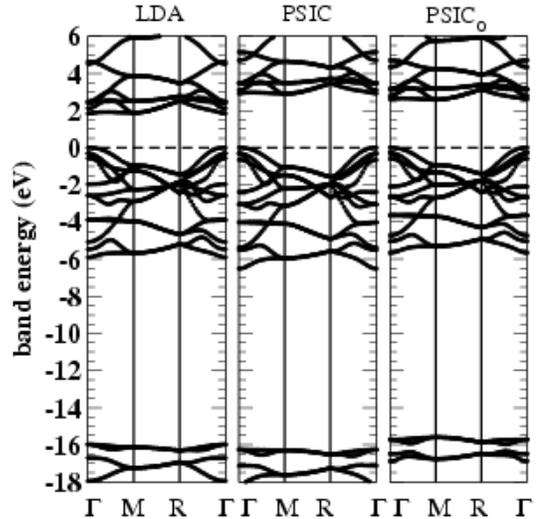}}
\caption{Band energies of TiO$_2$ (rutile) at experimental structure ($a$=4.59 \AA, $c$/$a$=0.664, u=0.305) calculated 
by LDA, VPSIC and VPSIC$_0$. K-points coordinates (units of $\pi$/$a$=$\pi$/$b$, $\pi$/$c$) are $\Gamma$=(0,0,0),
M=(1,1,0), R=(1,1,1).
\label{band_tio}}
\end{figure}

We start with the analysis of the band energies of TiO$_2$ in Fig.\ref{band_tio}, where results for VPSIC, VPSIC$_0$, 
and LDA are reported (for an unbiased comparison of the methods we consider the bands calculated at the same (experimental) structure). 
The corresponding band gap energy values are reported in Tab.\ref{tab_gap}. As expected, TiO$_2$ shows 
the energy gap between occupied O 2p valence and empty Ti 3d conduction bands. According to VPSIC, the minimum gap is 
indirect between $\Gamma$ at valence band top (VBT) and M (i.e. the BZ edge along [110]) at the conduction band bottom 
(CBB), while the direct gap is at $\Gamma$. The energy gap values are in satisfying agreement with the experiments,
while the LDA results present the well-known band gap underestimation of nearly 40\%, a typical LDA error bar for 
non-magnetic insulators. The PSIC$_0$, on the other hand, operates a partial (about 80\%) recovery of the correct energy 
gap over the LDA result. While this may be somewhat unsatisfying for the purpose of predicting accurate photoemission 
energies, it may be sufficient for performing structural optimization at a pace substantially similar to that of the 
LDA itself. Notice also that for what concern the (mainly O 2p) valence bandwidth, VPSIC and VPSIC$_0$ changes
the LDA value ($\sim$ 6.0 eV) in opposite directions ($\sim$ 6.5 eV for VPSIC, 5.7 eV for VPSIC$_0$); this illustrates the
fact that the VPSIC$_0$ is not simply a rescaled VPSIC.  

%%%%%%%%%%%%%%%%%%%%%%%%%%%%%%%%%%%%%%%%%%%%%%%%%%%%%%%%%%%%%%%%%%%%%%%%%%%%%%%%%%%%%%%%%%%
%%%%%%%%%%%%%%%%%%%%%%%%%%%%%%%%%%%%%%%%%%%%%%%%%%%%%%%%%%%%%%%%%%%%%%%%%%%%%%%%%%%%%%%%%%%%%%

It is worthy emphasizing that the capability of VPSIC (or VPSIC$_0$) to correct the LDA failure in case of non-magnetic 
oxides whose ground state is only residually affected by $d$-type orbitals, spurs from its all-orbital corrective character
(in particular from the O 2p band correction). In contrast, the GGA+U band gap of 2.2 eV (with U=3.4 eV applied to Ti 3d) 
calculated in Ref.\onlinecite{perevalov}, is only marginally larger than the respective GGA value of 1.9 eV. 
The necessity of applying a corrective U onto the O 2p orbital energies was indeed discussed in previous 
works\cite{nekrasov}. 

%%%%%%%%%%%%%%%%%%%%%%%%%%%%%%%%%%%%%%%%%%%%%%%%%%%%%%%%%%%%%%%%%%%%%%%%%%%%%%%%%%%%%%%%%%
\begin{figure}
\epsfxsize=5cm
\centerline{\includegraphics[clip,width=8.5cm]{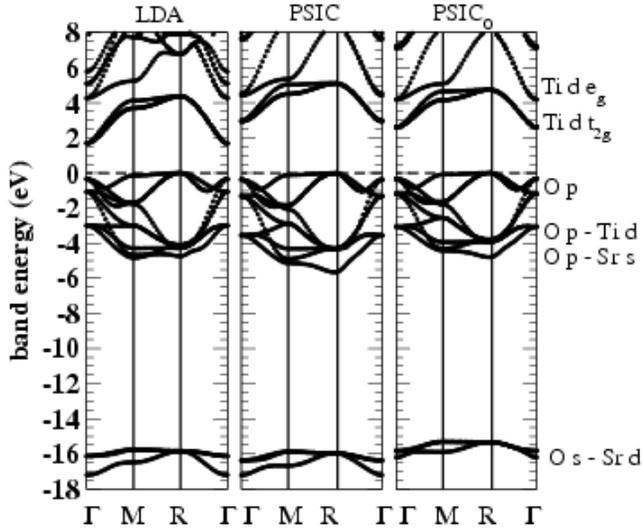}}
\caption{Band energies of cubic SrTiO$_3$ at experimental structure ($a$=3.92 \AA). K-points coordinates (units of $\pi$/$a$) are $\Gamma$=(0,0,0),
M=(1,1,0), R=(1,1,1).
\label{band_srtio}}
\end{figure}
%%%%%%%%%%%%%%%%%%%%%%%%%%%%%%%%%%%%%%%%%%%%%%%%%%%%%%%%%%%%%%%%%%%%%%%%%%%%%%%%%%%%%%%%%%%%%

The band energies for bulk cubic SrTiO$_3$ is reported in Figs.\ref{band_srtio}. The dominant (and eventually the 
secondary) orbital character for each group of bands is also reported in Figure. The energy gap opens between CBB of 
mainly Ti 3d character, with higher contributions from Sr 4d states, and VBT bands dominated by O 2p states. Wee can see 
in Tab.\ref{tab_gap} that the LDA band gap is only $\sim$ 55\% of the experimental direct (3.25 eV) and indirect (3.75 eV)
gap. The VPSIC, on the other hand, recovers most ($\sim$ 90\%) but not all the LDA discrepancy. This is in part 
attributable, rather than to a VPSIC insufficiency, to a much too low LDA value, due to our specific technicalities and 
used pseudopotentials. Indeed previous LDA determinations \cite{benthem, cappellini} gave 1.9 eV and 2.24 eV for indirect 
and direct band gap, and a "scissor" operator of 1.5 eV was employed in Ref.\onlinecite{benthem} to readjust band energies
with ellipsometry data. According to our calculations the VPSIC operate a "scissor" of about 1.3 eV with respect to the LDA
band gap. Notice also that while in general the action of VPSIC over the LDA bands may be largerly k-point dependent, for 
these wide-gap, highly ionic, non-magnetic oxides the LDA band dispersion is not dramatically modified, indeed, and the 
concept of scissor operator can be grossly justified "a posteriori". However, if the shape is not much changed, significant
differences are visible in the bandwith: consider e.g. the first unoccupied doublet of Ti 3d t$_{2g}$ character: in LDA it
spans about 2.5 eV, while in VPSIC or VPSIC$_0$ it is reduced to about 2 eV, in consequence of the enhanced charge 
localization (i.e. reduced p-d hybridization) which tipically follows from the removal of SI. On the other hand, the VPSIC
increases the LDA O 2p bandwidth as a consequence of the different spectral weight distribution (the top manifold is purely
O 2p, while the bottom (R point) shows a significant mixture of Ti 3d and Sr 3s states. The different effective SI energies
related to these different orbital characters eventually stretches the band bottom down to lower energies, thus resulting 
in an increase of bandwidth. In VPSIC$_0$ on the other hand the effective SI energies are fixed to their atomic 
counterparts, thus the effect over the LDA values is more similar to that of a rigid band shift.  

%%%%%%%%%%%%%%%%%%%%%%%%%%%%%%%%%%%%%%%%%%%%%%%%%%%%%%%%%%%%%%%%%%%%%%%%%%%%%
\begin{figure}
\centerline{\includegraphics[clip,width=8.5cm]{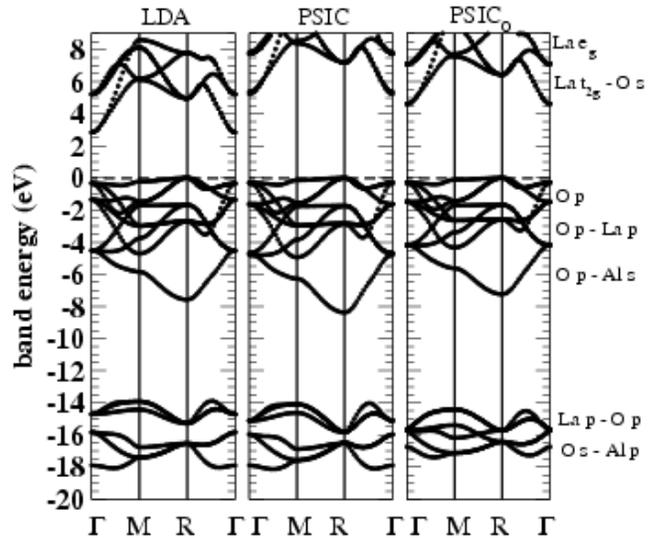}}
\caption{Band energies of cubic LaAlO$_3$ at experimental structure ($a$=3.82 \AA). K-points are the same as in 
Fig.\ref{band_srtio}.
\label{band_laalo}}
\end{figure}
%%%%%%%%%%%%%%%%%%%%%%%%%%%%%%%%%%%%%%%%%%%%%%%%%%%%%%%%%%%%%%%%%%%%%%%%%%%%%

Finally, in Fig. \ref{band_laalo} the band energies of cubic LaAlO$_3$ are reported. Now the energy gap is between O 2p 
valence bands and mainly Al s,p conduction bands. The absence of 3d states produces an observed gap (5.6 eV) much larger 
than in SrTiO$_3$, and again grossly underestimated in LDA ($\sim$56\% of the experimental value). This case is 
prototypical in demonstrating the necessity to repair the LDA unrespectively on the presence of 3d states. Again, VPSIC 
and VPSIC$_0$ works quite nicely in recovering a satisfying energy gap. However, as in case of SrTiO$_3$ they act in 
different ways for what concern the band width: VPSIC stretches the bottom of valence O 2p manifold down to lower 
energies, while in VPSIC$_0$ the same bands span an energy window even smaller than in LDA. For what concern the lower 
lying O s bands (between -16 eV and -18 eV) and La p (semicore) bands (around -14 eV), LDA and VPSIC give a quite similar 
description, while the two groups are overlapping according to VSIC$_0$.  

%%%%%%%%%%%%%%%%%%%%%%%%%%%%%%%%%%%%%%%%%%%%%%%%%%%%%%%%%%%%%%%%%%%%%%%%%%%%%%%%%%%%%%%%%%
\begin{table}
\caption{Lattice parameters (in \AA) calculated within LDA, VPSIC, and VPSIC$_0$, compared to the experimental values.
For SrTiO$_3$ and LaAlO$_3$ the cubic symmetry is assumed, while TiO$_2$ is in tetragonal (rutile) structure. 
For TiO$_2$ c/a is fixed to the experimental value 0.644, while the internal parameter u is calculated.
\label{tab_lattice}}
\centering\begin{tabular}{|lcccc|}
\hline
            & LDA & PSIC & PSIC$_0$ & Expt. \\
\hline\hline
 SrTiO$_3$ &   3.99  & 3.97  & 4.02   &   3.92     \\
 LaAlO$_3$ &   3.76 &  3.74  & 3.83   &   3.82     \\
 TiO$_2$ a  &   4.67   & 4.62    & 4.69   &  4.59   \\
 TiO$_2$ u  &   0.3021 & 0.3066  & 0.3066 &  0.3048  \\
\hline
\end{tabular}
\end{table}
%%%%%%%%%%%%%%%%%%%%%%%%%%%%%%%%%%%%%%%%%%%%%%%%%%%%%%%%%%%%%%%%%%%%%%%%%%%%%%%%%%%

\begin{figure}
\centerline{\includegraphics[clip,width=8.5cm]{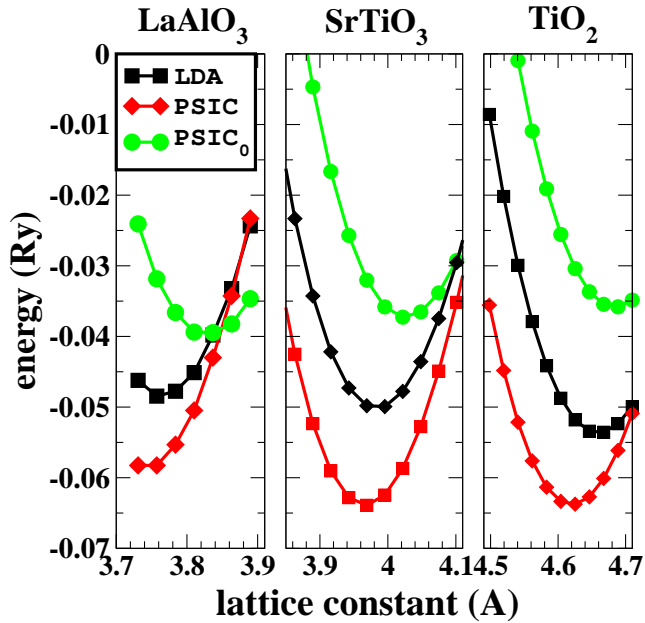}}
\caption{(Color on-line) Total energies as a function of lattice parameter for for cubic SrTiO$_3$, LaAlO$_3$ and 
tetragonal TiO$_2$, calculated by LDA, VPSIC, and VPSIC$_0$.
\label{struc}}
\end{figure}

In Fig.\ref{struc} we report total energies vs. lattice parameter for cubic SrTiO$_3$, LaAlO$_3$ and tetragonal TiO$_2$, 
calculated within LDA, VPSIC, and VPSIC$_0$ for the three examined oxides. The calculated equilibrium parameters 
are reported in Tab.\ref{tab_lattice}, in comparison with the experimental values. We can start the analysis considering
the reference LDA values: as mentioned, LDA is tipically satisfying for what concern structural properties, but the level 
of accuracy may vary depending on technicalities. Thus while LDA is known to generally underestimate by 1-2 \% the 
equilibrium lattice constant, we have values for SrTiO$_3$ and TiO$_2$ which overestimate the experiments by $\sim$ 1.5\%.
On the other hand, the lattice constant of LaAlO$_3$ is underestimated by little more than 1%. The VPSIC operates a 
systematic (0.5-1\%) reduction of the corresponding LDA values, thus ending up showing a very satifying accord 
with the experiment (within 1%) for SrTiO$_3$ and TiO$_2$, and a bit less satisfying (2\%) for LaAlO$_3$. The VPSIC$_0$ 
goes again in the opposite direction, as it increases sistematically the LDA values by 0.5\%-1\%.

%%%%%%%%%%%%%%%%%%%%%%%%%%%%%%%%%%%%%%%%%%%%%%%%%%%%%%%%%%%%%%%%%%%%%%%%%
%%%%%%%%%%%%%%%%%%%%%%%%%%%%%%%%%%%%%%%%%%%%%%%%%%%%%%%%%%%%%%%%%%%%%%%%

The different behaviour of VPSIC and VPSIC$_0$ can be easily linked back to what we have commented for the bands: the 
VPSIC tend to stretches the occupied valence band manifold (O 2p, O 2s) with respect of the LDA values, thus reducing the
effective screening and in turn the bond lenght. On the contrary, the VPSIC$_0$ shrinks the occupied band manifolds with 
respect to the LDA, thus causing an increase of effective screening and bond length. The shrinking of equilibrium volume
caused by the VPSIC was also reoprted for transition metal monoxides MnO and NiO\cite{tmo}. Clearly, this should not be 
intended as a 'universal' trend, as bandwidth and charge localization are case-dependent quantities, and as such are 
the VPSIC modifications over the LDA electronic properties. 

We can conclude this section emphasizing the overall good quality of the VPSIC predictions for the three examined
compounds, with lattice parameters and energy band gaps within 1-2\% and $\sim$10\% from the experiments, respectively.
While many more results will be necessary for a full assesment of the theory, what we have showed in this Section is
sufficient to encourage the use of VPSIC for the investigation of wide-gap oxide insulators.

%%%%%%%%%%%%%%%%%%%%%%%%%%%%%%%%%%%%%%%%%%%%%%%%%%%%%%%%%%%%%%%%%%%%%%%%%%%%%%
%%%%%%%%%%%%%%%%%%%%%%%%%%%%%%%%%%%%%%%%%%%%%%%%%%%%%%%%%%%%%%%%%%%%%%%%%%%%%%
%%%%%%%%%%%%%%%%%%%%%%%%%%%%%%%%%%%   TITANATES %%%%%%%%%%%%%%%%%%%%%%%%%%%%%%
%%%%%%%%%%%%%%%%%%%%%%%%%%%%%%%%%%%%%%%%%%%%%%%%%%%%%%%%%%%%%%%%%%%%%%%%%%%%%%
%%%%%%%%%%%%%%%%%%%%%%%%%%%%%%%%%%%%%%%%%%%%%%%%%%%%%%%%%%%%%%%%%%%%%%%%%%%%%%

\subsection{Magnetic titanates: YTiO$_3$, LaTiO$_3$}
\label{res_tita}

%%%%%%%%%%%%%%%%%%%%%%%%%%%%%%%%%%%%%%%%%%%%%%%%%%%%%%%%%%%%%%%%%%%%%%%%%%%%%%%%

Titanates characterized by the nominal Ti d$^1$ configuration rank among the most peculiar and intriguing magnetic 
perovskites. At variance with the more investigated classes of manganites and cuprates whose fundamental chemistry 
is governed by 3d e$_g$ states, in titanates the 3d valence states are 3d t$_{2g}$, thus orbitals not directly 
oriented towards the oxygens; this produces a much weaker p-d hybridization and crystal field splitting than in
e$_g$ systems. However, experiments show that the phenomenology of these systems may be crucially affected by 
small structural details. 

A nice illustration of this over-sensitive magnetostructural coupling is furnished by the 
compared study of YTiO$_3$ (YTO) and LaTiO$_3$ (LTO): both systems are Pnma perovskites, with relatively small 
Jahn-Teller (JT) distortions and large GdFeO-type octahedral rotations; the difference in cation size (with La bigger 
than Y) makes the amplitude of distortions and rotations sensibly wider in YTO than in LTO (in agreement with the well 
known space-filling criterion), in turn leading to quite a different magnetic behaviour: YTO is ferromagnetic 
(FM)\cite{khaliullin1,tsubota,nakao} with low Curie temperature T$_c$=30 K, sizeable band gap ($\sim$ 1 eV) and magnetic
moments M=0.8 $\mu_B$, in line with a Ti d$^1$ ionic configuration; LTO, on the other hand, is antiferromagnetic (AF) 
G-type\cite{mochizuki} with T$_N$=130 K, very small energy gap ($\sim$ 0.3 eV) and sensibly smaller magnetic moments 
($\sim$ 0.57 $\mu_B$)\cite{cwik}. A long-standing debate was ignited on the attempt to give an explaination to this much reduced 
magnetic moment and nearly isotropic spin-wave dispersion in LaTiO$_3$\cite{keimer}. It was pointed out that a single 
electron in the triple-degenerate t$_{2g}$ manifold can fluctuate giving rise to an exotic 'orbital liquid' 
state\cite{kiyama,khaliullin}. However this fashinating hypothesis is contrasted by a series of 
evidences\cite{cwik,craco,haverkort,hemberger,kiyama,mochizuki,lunkenheimer} that crystal field splitting is actually not 
small enough to keep the t$_{2g}$ degeneracy substantially unlifted, and instead a Jahn-Teller distorted, orbital-ordered 
state is realized in LTO, as well as in YTO.  

\begin{figure}
\centerline{\includegraphics[clip,width=7cm]{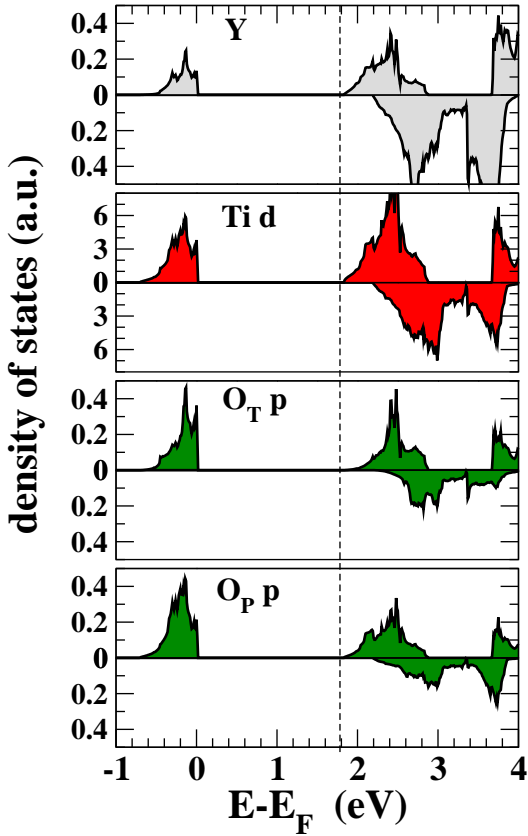}}
\caption{(Color on-line) Orbital-resolved DOS for FM YTO. For clarity only Ti 3d, Y 5d, and O 2p are showed 
(oxygens on-top and in plane with Ti are labelled O$_T$ and O$_P$, respectively). Notice that Y and O DOS are 
magnified by more than one order of magnitude with respect to the dominant Ti 3d DOS.
\label{dos_yto}}
\end{figure}

Needless to say, these issues stimulated a long series of attempts to describe the titanates by a variety of ab-initio 
approaches, including LSDA\cite{fujitani}, LDA+U\cite{sawada, streltsov}, and several LSDA+DMFT 
implementations\cite{pavarini,solovyev,nekrasov2}. While our description reproduces, at least in part, some of the previous findings,
one aspect makes our results very valuable: they capability to furnish a coherent rendition of structural and 
electronic properties of on a purely ab-initio ground, in the framework of the same theory, and without inclusion
of system-dependent parameters (e.g. U, J).
 
\begin{figure}
\centerline{\includegraphics[clip,width=9cm]{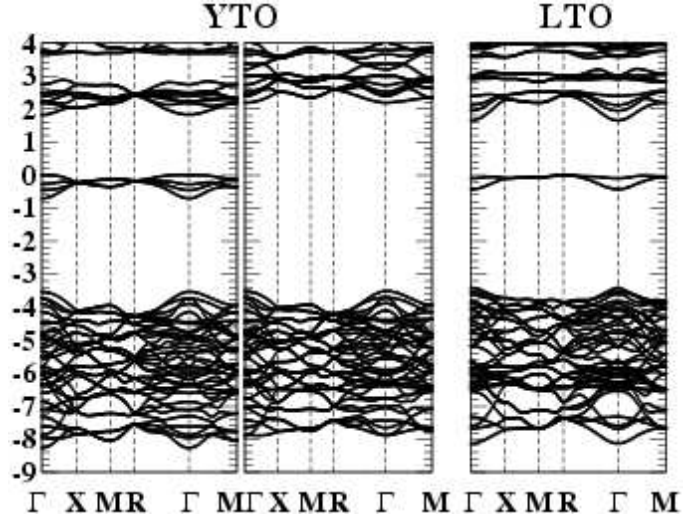}}
\caption{(Color on-line) Top: Band energies at x=1/4 (h=0.125) for (1$\times$1) PM phase (a1) and (2$\times$2) 
AF phase (b1). K-points coordinates (units of 1/a=1/b, 1/c with a, b, c unit-cell parameters) are X=[$\pi$/2,0,0], 
X�=[0,$\pi$/2,0], L=[$\pi$/2,$\pi$/2,0], M=[$\pi$/2,$\pi$/2,$\pi$/2]. Bottom: calculated FS for PM (a2) and AF (b2) 
phase.
\label{band_ti}}
\end{figure}

\begin{figure}
\centerline{\includegraphics[clip,width=7cm]{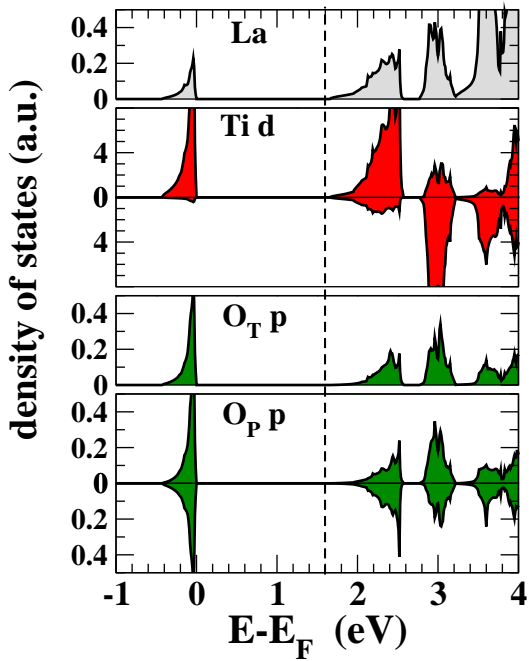}}
\caption{(Color on-line) Orbital-resolved DOS for AF G-type LTO (right). Orbital labels are the same as in Fig.\ref{dos_yto}.
La and O$_T$ states are spin-compensated, due to AFG symmetry. La and O DOS are magnified by more than one order of 
magnitude with respect to the dominant Ti 3d DOS.
\label{dos_lto}}
\end{figure}

\begin{figure}
\centerline{\includegraphics[clip,width=9cm]{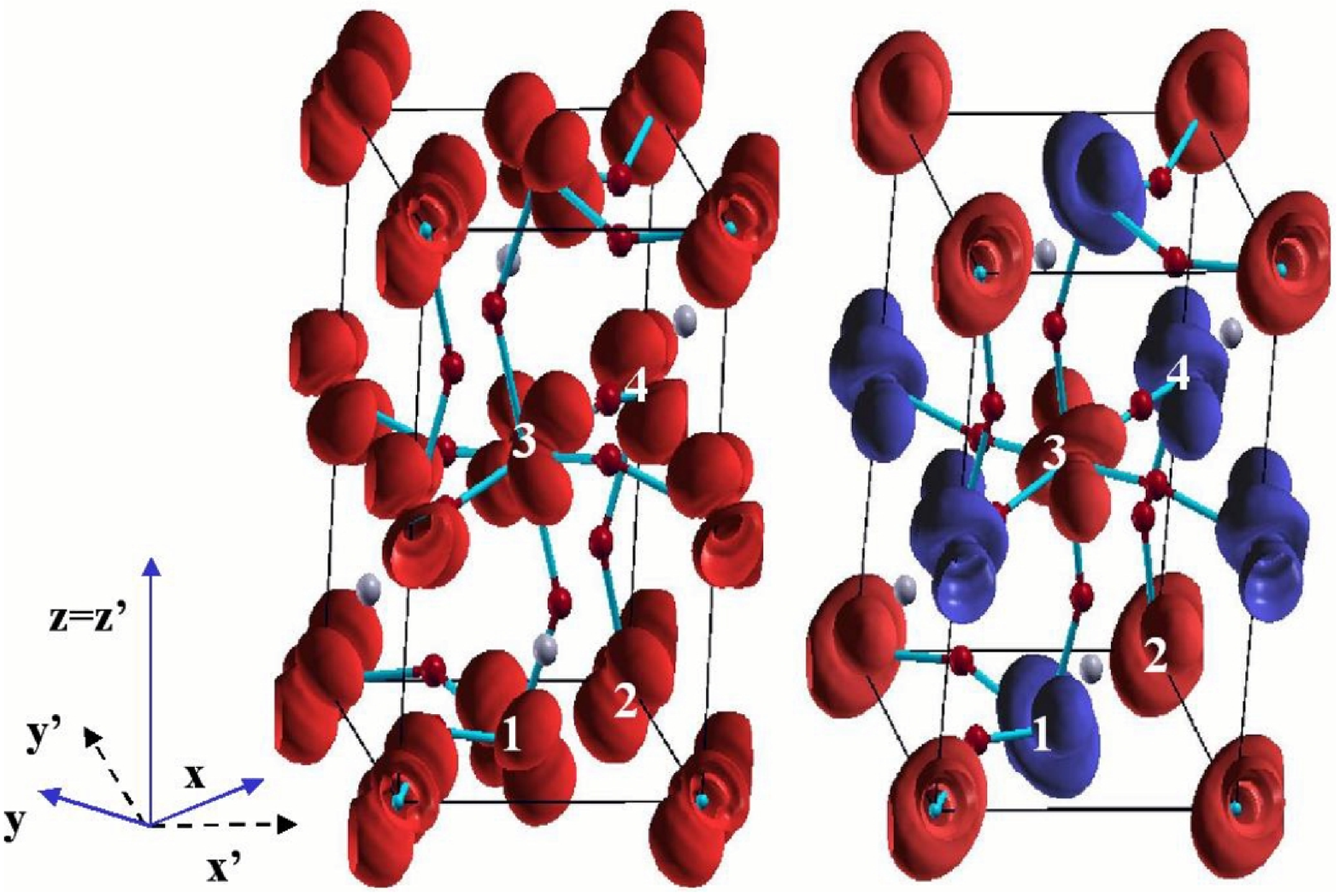}}
\caption{(Color on-line) Charge density isosurface $n_\pm$=$\pm$ 0.01 electrons/cm$^3$) of the upmost occupied state for 
FM YTO (left) and AF-G LTO (right). Red (light) and blue (dark) surfaces represent spin majority (+) and minority (-) 
contributions, respectively. On this scale only Ti d contributions are visible (oxygen contributes residually, see the 
DOS in Figs.\ref{dos_yto} and \ref{dos_lto}). Both YTO and LTO are
orbital-ordered, i.e. the four Ti atoms in the cell have same integrated charge but different orbital distribution 
(numbers connect each Ti with the corresponding 3d orbital decomposition reported in Tab.\ref{tab_occ}).
\label{magn_yto_lto}}
\end{figure}

We will proceed as following: we start illustrating the electronic properties of YTO, calculated at the experimental 
structure, as prototype of 'basic' t$_{2g}$ system, then we move to discuss the more peculiar LTO, highlighting the 
differences with respect to YTO, and finally we end up with the structural properties of both systems, which will furnish a 
rationale to their different behaviour. Notice that we do not include any comparison with local-density results in the 
analysis: LSDA does not reproduce the Mott-insulating behaviour for these systems and in fact predict an unphysical non 
magnetic, metallic electronic ground-state which is not meaningful for the purpose of confronting the VPSIC rendition. 

In Fig.\ref{dos_yto} the orbital-resolved DOS of YTO is showed. The occupied DOS have two major contributions: 
at VBT there is a $\sim$0.8 eV-wide fully spin-polarized DOS peak of Ti 3d states (residually hybridized with a 
small O 2p portion). Despite the nominal Ti$^{3+}$ d$^1$ configuration, a certain amount of Ti d-O p 
hybridization is clearly visible in Figure (notice however the different scale of Ti 3d and O 2p DOS: here the O 2p weight
is way smaller than, e.g. in manganites). It follows that the calculated static charges and magnetic moment differ 
considerably from their nominal values (for Ti we obtain $\sim$ 1.6 and $\sim$0.7 electrons for up and down-polarized 3d 
state, respectively, and M=0.92$\mu_B$, a bit larger than the observed 0.8$\mu_B$). Below the Ti 3d peak there is a 
broader, unpolarized DOS of O 2p character spanning the energy interval between -4 eV to -8 eV (not showed in Figure).
The CBB bands are also dominated by far and large by Ti 3d t$_{2g}$ s tates, residually hybridized with O 2p and Y 4d 
orbitals. Thus we can unquestionably categorize the system as a true Mott-Hubbard insulator, at variance with most
manganites or cuprates that are actually charge-transfer insulators or in the intermediate regime (we will be back later
on this important point).

In the band energy plot of FM YTO (Fig.\ref{band_ti}, left panel) we see four occupied bands (one for each Ti) 
separated from the 3d empty conduction bands by 1.8 eV. The fundamental gap only involves bands of majority manifold, 
and is direct at $\Gamma$. The CBB bands span a $\sim$ 1 eV wide energy interval.  According to our calculations, the 
sharp DOS peak at the valence top is a complex admixture of the five Ti 3d orbitals. To rationalize quantitatively the 
identity of this state, we have diagonalized the corresponding P$^{\sigma}_{mm'}$ density matrix in the 3d orbital 
subspace. The results are reported in Tab.\ref{tab_occ} for two coordinate systems: the orthorhombic 
Pnma $\sqrt{2}\times\sqrt{2}\times$2 (x',y',z'), and the conventional cubic(x,y,z), which differ by a 45$^o$ 
rotation\cite{rot} of the(x,y) plane. Focusing first on the Ti sited at (0,0,0) in the cubic reference system of YTO, 
we see that it shows a prevailing contributions of $|yz\rangle$ and ($|xy\rangle$) orbitals; however, 
not completely discardable e$_g$ contributions are present as well. The charge density isosurface plot 
(Fig.\ref{magn_yto_lto} left panel) shows that this state can be substantially expressed as 
$|\Psi_1\rangle$ $\sim$ 0.75 $|yz\rangle$ + 0.56$|xy\rangle$. Also evident is the resulting orbital ordering: co-planar 
states shows an alternance of dominant 
$|yz\rangle$ and $|xz\rangle$ contributions, plus a change of sing for $|xy\rangle$ which makes the lobes of 
$|yz\rangle$ (or $|xz\rangle$) leaning back and forth towards the (x,y) plane 
(e.g. $|\Psi_2\rangle$ $\sim$ 0.75 $|xz\rangle$ - 0.56$|xy\rangle$). On the other hand, states aligned along z only differ
by the alternance of $|xy\rangle$ sign, thus $|\Psi_3\rangle$ $\sim$ 0.75 $|yz\rangle$ - 0.56$|xy\rangle$, 
$|\Psi_4\rangle$ $\sim$ 0.75 $|xz\rangle$ + 0.56$|xy\rangle$. Our results are in excellent agreement with the finding of 
linear dichroism x-Ray absorption measurement\cite{iga} which gives 0.8 and 0.6 for the coefficients of the two most 
occupied t$_{2g}$ orbitals.\cite{note_iga}, and with LDA+DMFT results\onlinecite{pavarini} which obtain 
0.78 and 0.62, respectively.

\begin{table}
\begin{center}
\caption{3d orbital decomosition of the four occupied states (one for each Ti) at VBT of YTO and LTO. Coordinates 
(x',y',z') and (x,y,z) refers to orthorhombic and conventional cubic cartesian axes, respectively, as indicated in 
Fig.\ref{magn_yto_lto}. 
\label{tab_occ}}
\begin{tabular}{lccccc}
		& $|x'y'\rangle$      &  $|x'z'\rangle$      &   $|y'z'\rangle$     & $|z'^2\rangle$    &  $|x'^2-y'^2\rangle$     \\
\hline
YTO           &           &            &            &           &                  \\
Ti 1          &  0.11     &  0.48      &   0.58     &  0.33     &    0.56          \\
Ti 2          &  0.11     &  0.48      &  -0.58     & -0.33     &   -0.56          \\
Ti 3          & -0.11     &  0.48      &   0.58     & -0.33     &   -0.56           \\
Ti 4          & -0.11     &  0.48      &  -0.58     &  0.33     &    0.56          \\
\hline 
LTO           &           &            &            &           &                  \\
Ti 1          &  0.02     &  0.15      &   0.78     &  0.08     &    0.60          \\
Ti 2          &  0.02     &  0.15      &  -0.78     & -0.08     &   -0.60          \\
Ti 3          & -0.02     &  0.15      &   0.78     & -0.08     &   -0.60          \\
Ti 4          & -0.02     &  0.15      &  -0.78     &  0.08     &    0.60          \\
\hline
\hline
		& $|xy\rangle$      &  $|xz\rangle$      &   $|yz\rangle$     & $|z^2\rangle$    &  $|x^2-y^2\rangle$     \\
\hline
YTO          &         &              &          &          &           \\
Ti 1         & 0.56    &  -0.07       &  0.75    &  0.33    &  0.11       \\
Ti 2         &-0.56    &   0.75       & -0.07    & -0.33    &  0.11     \\ 
Ti 3         &-0.56    &  -0.07       &  0.75    & -0.33    & -0.11       \\
Ti 4         & 0.56    &   0.75       & -0.07    &  0.33    & -0.11      \\
\hline		
LTO          &         &              &          &          &            \\
Ti 1         & 0.60    &  -0.45       &  0.66    &  0.08    &   0.02      \\
Ti 2         &-0.60    &   0.66       & -0.45    & -0.08    &   0.02       \\
Ti 3         &-0.60    &  -0.45       &  0.66    & -0.08    &  -0.02        \\
Ti 4         & 0.60    &   0.66       & -0.45    &  0.08    &  -0.02       \\
\hline
\end{tabular}
\end{center}
\end{table}

Now we move to analyze the results for LTO; remarkable differences from YTO emerge from the calculated DOS 
(Fig.\ref{dos_lto}) and band structure (\ref{band_ti}): the fundamental band gap is a bit smaller for LTO but still
quite sizeable (1.6 eV); furthermore, the latter presents a band flatness which is stronger than in YTO: the occupied 
3d states at VBT now span a much narrower energy range (0.4 eV instead of 0.8 eV), and the hybridization with the oxygens
is more residual, although still well visible. Even the conductions bands in LTO appear flatter, and in fact they are 
separated in two groups by a gap of 0.2 eV. The magnetic moment is 0.89 $\mu_B$, similar to YTO.

The difference with YTO is also well spelled by the analysis of the diagonalized density matrix in Table \ref{tab_occ}. 
Looking at the cubic reference system, at variance with YTO we see that the e$_g$ contribution is now almost vanishing, 
and the occupied states are almost purely t$_{2g}$. Moreover, the diversification of the t$_{2g}$ occupancies is much 
reduced with respect to YTO as a results of the smaller rotations, and indeed this state approximately resembles 
cygar-shaped [111]-directed lobes resulting by the near even t$_{2g}$ combination, as confirmed by the corresponding 
charge density isosurface plot in Fig.\ref{magn_yto_lto}, right panel (notice that if t$_{2g}$ coefficients were exactly the 
same, $\Psi_1$ and $\Psi_2$, as well as $\Psi_3$ and $\Psi_4$, would have been identical, and the resulting cygars in each
plane parallel to each other, pointing all along [111]). 

While the orbital charge distribution in YTO and LTO is so different, and causes much of their macroscopic differences, 
the relative ordering is the same: even for LTO, in the plane there is perfect alternance (i.e. chessboard-like 
order) of leading $|xz\rangle$ and $|yz\rangle$ contributions (this is less evident than in YTO since the t$_{2g}$ 
coefficients are not as different as in YTO), plus a sing alternance for $|xy\rangle$. Along z only the sing alternance 
occurs. Our calculated t$_{2g}$ coefficients are remarkably close to the values (0.56, 0.45, 0.69) measured by NMR spectra
in Ref.\onlinecite{kiyama}, as well as those calculated by a model Hamiltonian (0.6, 0.39, 0.69) in 
Ref.\onlinecite{mochizuki}. 

The observed magnetic ground-state is correctly predicted for both systems: for YTO we found the FM energy lower than AF-G
and AF-C phases by 10.1 meV/f.u. and 8.3 meV/f.u., respectively (in agreement with previous LDA+U results\cite{sawada}
with U-J=3.2 eV). For LTO, on the other hand, we obtain the AF-G phase lower than FM and AF-A phases by 15.2 meV/f.u. and 
10.05 meV/f.u., respectively. Fitting the energies on a 2-parameter nearest-neighbor Heisenberg Hamiltonian:
 
\begin{eqnarray}
H=-{1\over 2} \sum_i \left[ J_{pl} (\hat{S}_i\cdot\hat{S}_{i+x}+\hat{S}_i\cdot\hat{S}_{i+y})+
J_z \hat{S}_i\cdot\hat{S}_{i+z}\right]
\end{eqnarray}

where $i+x$, $i+y$, and $i+z$ indicate nearest neighbors of $i$ in the x,y, and z directions, respectively, we obtain 
J$_{pl}$= 4.15 meV and J$_z$=1.8 meV for planar and orthoghonal exchange interaction parameters in YTO, respectively;
J$_{pl}$= -5.02 meV and J$_z$=-5.03 meV for the same quantities in LTO. These results nicely confirm the conjectures
derived by the analysis of the orbital ordering: while a remarkable anisotropy is present in YTO, LTO is substantially 
isotropic.  

\begin{figure}
\centerline{\includegraphics[clip,width=9cm]{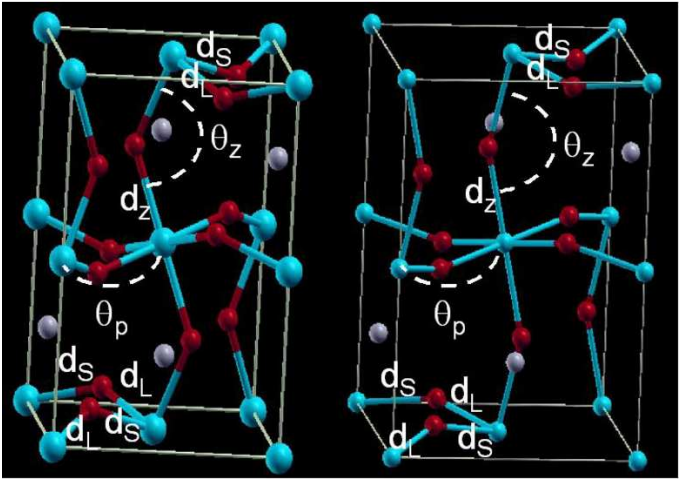}}
\caption{(Color on-line) Pnma structure of YTO (left) and LTO (right). Cell parameters are fixed to experimental values, while atomic positions
were relaxed according to VPSIC. Labels indicate Ti-O-Ti angles and Ti-O distances in plane ($\theta_p$, d$_p$) and along z ($\theta_z$, d$_z$).
Results are reported in Tab.\ref{tab_struc_yto_lto}.
\label{struc_yto_lto}}
\end{figure}

\begin{table}
\begin{center}
\caption{Atomic positions in crystal coordinates ($x/a$, $x/b$, $x/c$), and main structural parameters (Ti-O-Ti angles in plane ($\theta_p$) and 
along z ($\theta_z$), Ti-O distances along z (d$_z$) and in-plane (the shorter d$_S$ and the longer d$_L$) bonds) for Pnma YTiO$_3$ and LaTiO$_3$ 
calculated by VPSIC, in comparison with the experimental data (in parenthesys). Cell structures are fixed to the experimental values $a$=5.316 \AA, 
$b$=5.679 \AA, $c$=7.611 \AA $\>$ for YTO, $a$=5.640 \AA, $b$=5.584 \AA, $c$=7.896 \AA $\>$ for 
LTO\cite{cwik}. 
\label{tab_struc_yto_lto}}
\begin{tabular}{lccc} 
\hline
		& x/a     &y/b   &z/c		\\
\hline
\hline
Y 		& 0.478 (0.479)   & 0.073 (0.073)    & 1/4	\\		
Ti 		& 0  	            & 0                & 1/2      \\		
O$_{I}$	& -0.139 (-0.121) &-0.063 (-0.042)   & 1/4	\\		
O$_{II}$    &  0.307 (0.309)  & 0.184 (0.190)    & 0.067 (0.058) \\		
\hline
La 		& 0.491(0.493)    & 0.053(0.043)     & 1/4	\\		
Ti 		& 0  	            & 0                & 1/2      \\		
O$_{I}$	& -0.080(-0.081)  & -0.008(-0.007)   & 1/4	\\		
O$_{II}$    & 0.0288(0.291)   & 0.204(0.206)     & 0.042 (0.043)  \\		
\hline
\hline
            &   d$_S$   &   d$_L$      &   d$_z$       \\
\hline
YTO         &   2.0(2.02)  &   2.13(2.08)  &  2.07(2.02)  \\
LTO         &   2.02(2.03) &   2.06(2.05)  &  2.02(2.03)  \\
\hline
\end{tabular}
\begin{tabular}{lcc} 
\hline
            &   $\theta_p$               &   $\theta_z $    \\
\hline
YTO		&  140.41$^o$ (143.62$^o$)   &  133.30$^o$ (140.35$^o$)    \\
LTO		&  153.82$^o$ (152.93$^o$)   &  154.30$^o$ (153.75$^o$)     \\
\hline
\end{tabular}
\end{center}
\end{table}

Table \ref{struc_yto_lto} shows experimental and VPSIC-calculated atomic coordinates and the most important structural 
parameters, i.e. Ti-O-Ti angles ($\theta$), and Ti-O distances, indicated in  Fig.\ref{struc_yto_lto}. In plane there are two 
types of Ti-O bonds, long (d$_L$) and short (d$_S$), which alternate along both x and y directions 
(see Fig.\ref{struc_yto_lto}),
while along $z$ there is only one d$_z$ $\sim$ d$_S$. These values easily rationalize the chessboard-like Ti d ordering: 
on each Ti the occupied state prefers to lie along the longer Ti-O bond (thus alternatively $|xz\rangle$ and $|xy\rangle$ 
for d$_L$ parallel to $x$, or $|yz\rangle$ and $|xy\rangle$ for d$_L$ parallel to $y$). For YTO the difference between 
d$_L$ and d$_S$ is quite sizeable, and give rise to a very pronounced ordering, as seen in the analysis of the charge 
density. For LTO the d$_L$ and d$_S$ difference is much reduced, and so is the planar chessboard ordering, indeed. 
Notice that for both materials the JT-type character of the structural distortions is quite residual, i.e. properties 
along $x$, $y$, and $z$ are, on average, almost the same (especially for LTO). The GdFeO-type tiltings and rotations, 
on the other hand, are quite sizable and represent the major factor determining the observed structures and the 
consequent splitting of the t$_{2g}$ triplet state. Finally, VPSIC-calculated structure is satisfactorily close to the 
experimental determination for both LTO and YTO (although for the latter oxygen rotations are a bit overemphasized along 
$z$ axis).

In summary, our VPSIC results furnish a coherent guideline to understand (at least part of) the differences between YTO 
and LTO, and correctly describes the different magnetic ordering of the two systems: The bigger GdFeO-type distortions of 
YTO produce crucial differences in electronic and magnetic properties, as evidenced by our results: a) larger Ti 3d-O 2p 
and t$_{2g}$-e$_{g}$ mixing; b) crucially different charge density distrubution around Ti; c) an increase by factor 
$\sim$2 of the occupied 3d state bandwidth. The wider rotations, in particular, destibilize the AF superexchange coupling 
which prevails in a purely d$^1$ t$_{2g}$ unrotated Pnma environment.  

Notice that the above considerations are completely reverted for doped manganites, whose chemistry is governed by 
e$_g$: in that case cubic symmetry and absence of octahedral rotations works in favor of e$_g$-p hybdridization. 
In titanates, on the other hand, absence of octahedral rotations means vanishing p-d hybridization, pure $_{2g}$ charge 
character, and minimal $_{2g}$ bandwidth.

It remains to explain the large difference between our calculated and the measured energy gap. This actually occurs by
construction: our VBT and CBB band energies represent removal and additional energies, and their difference estimates
the on-site Coulomb energy U, whereas the lowest excitation measured for these true Mott-Hubbard insulators is an 
intra-site excitation energy which of course does not include U. The attempt to argument a presumed smallness of U on 
the basis of the very tiny energy gap of LaTiO$_3$\cite{lunkenheimer} is a misinterpretation. In fact, 
according to our band structure U $\sim$3 eV, as expected for a system of this kind. It is also not very proper the
strategy carried out in several works of estimating the excitation energy as LDA-calculated t$_{2g}$ (average) band 
splitting: this is justified by the fact that in the limit of vanishing U (i.e. delocalized electrons) excitation and 
additional/removal energies go back to be the same quantities; however we must keep in mind that here the vanishing of 
U is an artifact of the LDA, not a true feature of the titanate. A more rigorous strategy to evaluate the lowest 
excitation energy is suggested in ref.\onlinecite{streltsov} where a suited Hamiltonian for the excited state is 
constructed in such a way to project out the electronic ground state. Here we do not pursue
this route which overcomes the capability our actual methodological setting, and leave to future developments the 
investigation of the crystal-field splitting and orbital-liquid state in LTO by VPSIC. However, we emphasize that the 
rationalization of the FM vs. AF G-type competition is correctly described already at the level of our ground-state 
results.

%%%%%%%%%%%%%%%%%%%%%%%%%%%%%%%%%%%%%%%%%%%%%%%%%%%%%%%%%%%%%%%%%%%%%%%%%%%%%%%%%%%%%%%%%%%%%%%%%
%%%%%%%%%%%%%%%%%%%%%%%%%%%%%%%%%%%%%%%%%%%%%%%%%   MANGANITES   %%%%%%%%%%%%%%%%%%%%%%%%%%%%%%%%
%%%%%%%%%%%%%%%%%%%%%%%%%%%%%%%%%%%%%%%%%%%%%%%%%%%%%%%%%%%%%%%%%%%%%%%%%%%%%%%%%%%%%%%%%%%%%%%%%
%%%%%%%%%%%%%%%%%%%%%%%%%%%%%%%%%%%%%%%%%%%%%%%%%%%%%%%%%%%%%%%%%%%%%%%%%%%%%%%%%%%%%%%%%%%%%%%%%

\subsection{Magnetic manganites: CaMnO$_3$}
\label{res_manga}

CaMnO$_3$ (CMO) is a prototype AF G-type insulator. The nominal Mn$^{4+}$ d$^3$ configuration triggers 
AF superexchange coupling via the fully polarized majority t$_{2g}$ orbitals, and AF semi-covalent exchange
interactions through the empty e$_g$ states. The t$_{2g}$ spherical charge distribution favours a robust
centrosymmetric octahedral structure, and the near complete absence of rotations leaves the systems substantially 
cubic (a small Pnma distortion is actually observed, but it will not be considered here, since it
is immaterial for magnetic and electronic properties). According to Goodenough-Kanamori rules\cite{goodenough}, 
spin coupling is expected to be AF and isotropic (G-type). While this is indeed verified by a series of 
experiments and calculations, the determination of the electronic and magnetic properties in detail is less
clear and some discrepancies between the interpretation of photoemission data and the band energies obtained
by standard local functionals make the system an ideal case for testing new methods. Indeed, from the theoretical 
side CMO was studied in the past using LDA\cite{pickett, satpathy, cardoso, dasgupta, fh}, GGA\cite{bhatta}, 
LDA+U \cite{satpathy}, GGA+U\cite{luo}, unrestricted Hartree-Fock (HF)\cite{nicastro,nicastro2}, configuration 
interaction (CI)\cite{nicastro2}, and model (Hubbard) Hamiltonian\cite{meskine}, while experimentally a number of 
optical\cite{park,jung,zampieri,zeng} and transport\cite{briatico} measurements has been carried out.

\begin{figure}
\centerline{\includegraphics[clip,width=8.5cm]{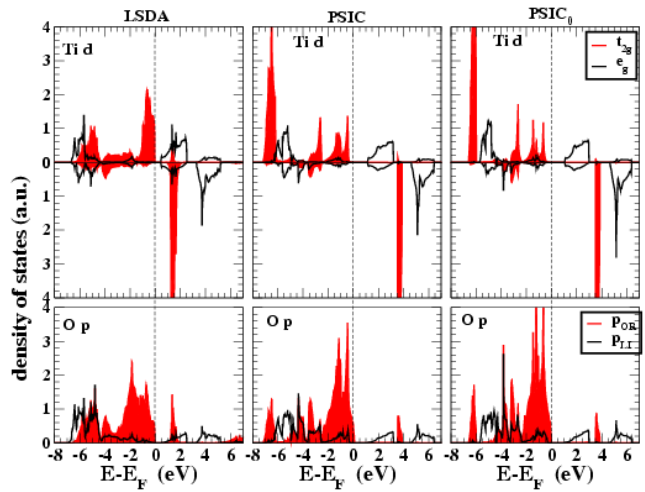}}
\caption{(Color on-line) Density of states of the most important orbitals (Mn d and O p) of AF-G CMO calculated within
LDA, VPSIC, and VPSIC$_0$. Light (red) shadowed areas are for Mn t$_{2g}$ and O p orthoghonal orbitals; solid black lines
for Mn e$_g$ and O p ligand orbitals.
\label{dos_cmo}}
\end{figure}
 
\begin{figure}
\centerline{\includegraphics[clip,width=8.5cm]{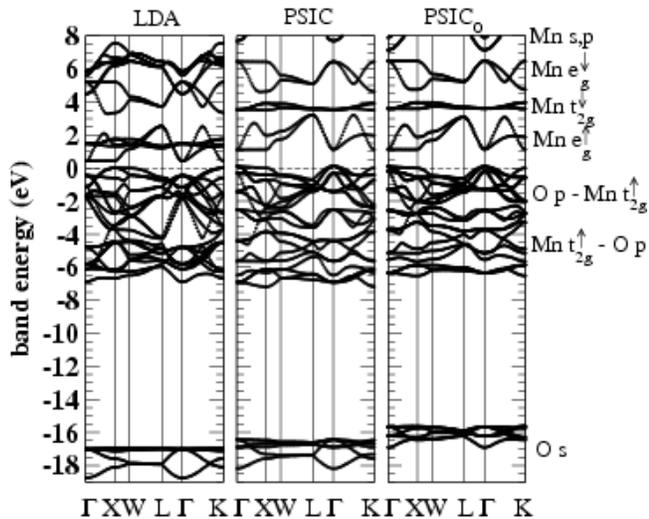}}
\caption{ Band energies of AF-G type CMO in cubic FCC symmetry, calculated by LSDA, VPSIC, and VPSIC$_0$. 
K-points coordinates (units of 2$\pi$/a) are $\Gamma$=[0,0,0], X=[1,0,0], W=[1,1/2,0], L=[1/2,1/2,1/2], K=[1,1,0].
Dominant and secondary orbital character for each group of bands is also indicated.
\label{band_cmo}}
\end{figure}

The general characteristics can be illustrated with the help of our calculated DOS (Fig.\ref{dos_cmo}) and band energies
(Fig.\ref{band_cmo}) for the observed AF G-type phase at experimental lattice parameter. All theories (LDA, VPSIC, and 
VPSIC$_0$) describe the system as insulator, with a $\sim$7 eV wide valence band manifold of mixed O p and majority Mn 
t$_{2g}$ states. Very importantly, at variance with the nominal configuration, a consistent amout of filled e$_g$ states 
is present, in fact, in all the three calculated DOS. Below the p-d valence bands a narrow O 2s band manifold lies
at about 18 eV from the VBT. Moving up in the energy above the fundamental gap we find distinct groups of majority $e_g$,
minority $t_{2g}$, and minority $e_g$ states.

\begin{table}
\begin{center}
\caption{Equilibrium lattice parameter a$_0$ (in \AA) for AF-G and FM phases and exchange interactions J (in meV) of cubic CaMnO$_3$ calculated by 
LSDA, VPSIC, and VPSIC$_0$ (experimental values are reported for comparison). J$_{eq}$ and J$_{ex}$ are values calculated for equilibrium and experimental 
a$_0$, respectively.
\label{tab_camno}}
\begin{tabular}{lcccc} 
\hline
		        & LSDA     &  VPSIC  & VPSIC$_0$ & expt. \\
\hline
\hline
a$_0$ (AF-G)	  & 3.75     & 3.74   & 3.83	   & 3.734 \\		
a$_0$ (FM)		  & 3.77  	 & 3.76   & 3.88     &       \\
\hline
\hline
J$_{ex}$            & -37.0	 & -6.1   &	 +6.5	    & -6.6\cite{rw}     \\			
J$_{eq}$            & -35.3    & -5.7   &  +27.3    &                   \\		
\hline
\end{tabular}
\end{center}
\end{table}

Looking deeper at the DOS important differences appear in the LSDA and VPSIC/VPSIC$_0$ description: for both the p-d valence
manifold shows a double peaked structure, in agreement with X-Ray (XPS) and ultraviolet (UPS) photoemission\cite{jung}, 
but the orbital character of the peaks is different: in LSDA the big part of t$_{2g}$ spectral weight is right below the 
VBT, while the tail region from -5 eV to -7 eV is mainly O 2p. The VPSIC/VPSIC$_0$ on the other hand recovers a spectral 
redistribution more in line with the observations, with prevalently O 2p states near VBT, and the highest peak of 
t$_{2g}$ states at the bottom of the valence bands. The LSDA failure is clearly related to the insufficient t$_{2g}$ spin
splitting, which amounts to a mere 2.5 eV and leaves the majority t$_{2g}$ much too high in the energy. In VPSIC the 
t$_{2g}$ splitting increases up to about 9 eV, which is consistent with the estimated value of U\cite{meskine}. Notice 
that a single energy parameter is not enough, however, to properly locate the t$_{2g}$ states, since a consistent portions
of those is also present in the 4 eV-wide region below the VBT, where the p-d hybridization is strong. 

In LSDA we obtain an energy gap of 0.42 eV. Looking at the band picture, the VBT runs flat between X and W, while the CBB
e$_g$ is flat between $\Gamma$ and X, in agreement with previous LDA calculations (see e.g.\onlinecite{dasgupta});
in VPSIC the gap is 1.01 eV, and both VBT and CBB are flat between $\Gamma$ and X. Above the energy gap, LSDA describes the 
2 eV-wide majority e$_g$ bands overlapped with the very narrow minority t$_{2g}$ peak, whereas in VPSIC/VPSIC$_0$ the latter
lies about 2 eV above the centroid of the e$_g$. While we could not find in literature a clear determination of the band 
gap value, interband transitions extracted from photoemission\cite{zampieri} and optical conductivity 
measurements\cite{jung} seem to be very consistent with the VPSIC calculation. Specifically, the distance between O 2p and
majority $e_g$ peaks ($\sim$3 eV) and between O 2p and minority $e_g$ peak ($\sim$ 6.5 eV) compares excellently with the 
respective values 3.07 eV and 6.49 eV extracted by Lorentz oscillator fitting of conductivity spectra reported in 
Ref.\onlinecite{jung}. Also quite consistent, albeit with a bit larger value (3.7 eV) for the O 2p-Mn e$_g$ transition, 
is the valence band spectra deduced in Ref.\onlinecite{zampieri} by fitting a CI cluster model to XPS data.

\begin{figure}
\centerline{\includegraphics[clip,width=6.5cm]{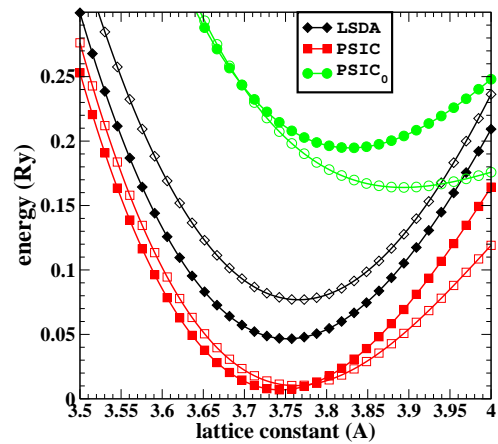}}
\caption{(Color on-line) Total energies per cell vs. lattice parameter calculated by LSDA, VPSIC, and VPSIC$_0$ for
AF-G (solid symbols) and FM (open symbols) ordering. Notice that differences in total energy due to different methods have 
no physical meaning and have been arbitrarily translated in the Figure for the sake of clarity.
\label{struc_cmo}}
\end{figure}

Now we move to examine structural and magnetic properties. In Fig.\ref{struc_cmo} total energies as a functon of lattice 
parameter for AF-G and FM phases are reported, calculated with LSDA, VPSIC, and VPSIC$_0$. Values of the equilibrium lattice
are reported in Tab.\ref{tab_camno}. The trend is similar to that seen in Section \ref{res_para} for wide-gap oxides: VPSIC and
VPSIC$_0$ respectively reduces and expands the volume with respect to the LSDA value. Here however, the correction of VPSIC 
is very tiny, and both LSDA and VPSIC gives lattice constant in very good accord (within 1\%) with the experimental value. 
On the other hand, the VPSIC$_0$ is much less satisfying, and the predicted equilibrum lattice overestimate the experiment
by $\sim$2.5\%. For what concern the difference between AF-G and FM energies, LSDA is known to exaggerate the contribution 
of AF superexchange due to the excessive t$_{2g}$-p hybridization in the region near VBT (discussed in Fig.\ref{dos_cmo}),
thu predicting a strong AF-G stability (in agreement with previous LSDA calculations\cite{pickett, solovyev} through the 
whole examined range of lattice values. The VPSIC, on the other hand, describes a much tighter competition: while at 
equilibrium the AF-G phase is stable, a moderate lattice stretching (about 1\% tensile strain) is sufficient to reverse 
the ordering and stabilize the FM ground state. Finally, the VPSIC$_0$ apparently performs very poorly for magnetic 
coupling: it completely reverses the LSDA description, and predicts the FM phase as robustly stable in a wide lattice 
range.

\begin{figure}
\epsfxsize=5cm
\centerline{\includegraphics[clip,width=6.5cm]{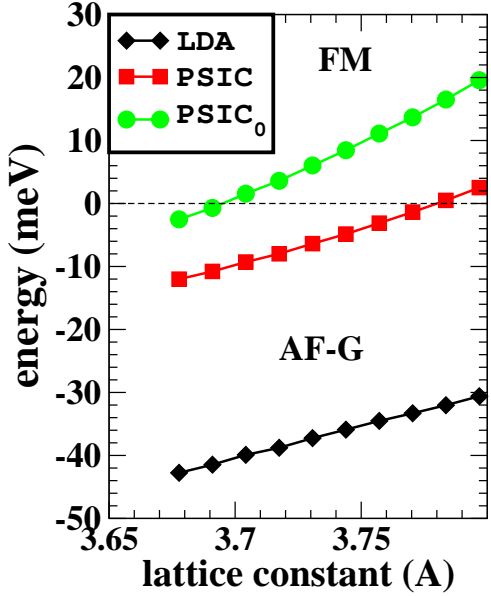}}
\caption{(Color on-line) Exchange-interaction parameter J calculated in LSDA, VPSIC, and VPSIC$_0$; see text for the exact 
definition of J.
\label{j_cmo}}
\end{figure}

The magnetic competition can be better appreciated in terms of exchange-interaction parameters J, plotted in 
Fig.\ref{j_cmo} (for the sake of comparison we have adopted the same definition of J given in Ref.\onlinecite{nicastro2}, 
based on the single-parameter Heisenberg Hamiltonian $H=-J\sum_{ij}\hat{e}_i\cdot \hat{e}_j$, where $\hat{e}_i$ is the 
versor of the $i$-site spin direction). Values calculated at equilibrium ($J_{eq}$) and experimental ($J_{ex}$) structure 
are reported in Table \ref{tab_camno}. We can remark the excellent agreement of the VPSIC value (-6.1 meV) with J=-6.6 meV 
extracted from the diagrammatic Rushbrook-Wood formula\cite{rw} for the magnetic susceptibility corresponding to the 
experimental Neel temperature T$_N$=130 K. The VPSIC also compares fairly well with those calculated by CI (8.1 meV) and 
HF (10.7 meV) in Ref.\onlinecite{nicastro2}. On the other hand, both LSDA and VPSIC$_0$ largely deviates from these 
estimates, albeit in opposite directions.

\begin{figure}
\epsfxsize=5cm
\centerline{\includegraphics[clip,width=8cm]{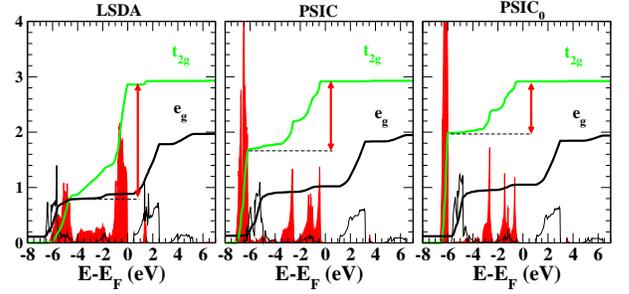}}
\caption{(Color on-line) Integrated DOS of t$_{2g}$ (green) and e$_g$ (black) majority orbitals, normalized to their 
respective degeneracies 3 and 2. The DOS are also reported for clarity. Lenght of red arrows indicate the amount of 
t$_{2g}$ charge without the lowest t$_{2g}$ peak. 
\label{dos2_cmo}}
\end{figure}

It is interesting speculating on the remarkably different magnetic behaviour described by the three methods. This mainly 
reflects the difference in the t$_{2g}$ spectral weight seen in Fig.\ref{dos_cmo}, specifically the substantial DOS shift
from the top to the bottom of the p-d band manifold when moving from LSDA to VPSIC and to VPSIC$_0$. As a quantification
of this effect, in Fig.\ref{dos2_cmo} we report the integrated charge of majority t$_{2g}$ and e$_g$ bands as described by
the three methods: all methods describe a filled (i.e. integrated to 3) t$_{2g}$ state, but if we evaluate the amount of
t$_{2g}$ charge located in the upper part of the valence band region which does not include the towering lower-end peak
and mainly hybridizes with oxygens (quantified in Fig.\ref{dos2_cmo} by the vertical red arrows) we find that this charge
in LSDA is about 73\%, in VPSIC just 43\%, and finally in VPSIC$_0$ a mere 33\% of the whole majority t$_{2g}$ charge. 
Our interpretation is the following: the lower-end t$_{2g}$ peak is too low in energy and too little hybridized with 
oxygens to give a meaningful superexchange contribution (through t$_{2g}$-p$_{OR}$ $\pi$-type bonding). Thus while the 
t$_{2g}$ spectral weight in transferred to the lower end of the valence band, the e$_g$ charge distribution remains 
substantially similar across the three renditions, and its FM  contribution takes place and even becomes dominant 
according to VPSIC$_0$. Notice that the e$_g$ charges integrates to a remarkable 1/2 electron per orbital at the VBT 
(in agreement with previous calculations\cite{luo}) thus it represent a prototypical case of FM coupling 
(via e$_g$-p$_{LI}$ hybridization) according to Goodenough-Kanamori rules\cite{goodenough}. 

In summary, the VPSIC seems to furnish a very consistent description of CMO for what concern electronic, structural, and 
magnetic properties. Values calculated in VPSIC compares well with the experiments and with results of other beyond-local 
approaches, when available, correcting the most obvious deficiencies of LSDA. On the other hand the VPSIC$_0$, while 
furnishing a band spectrum substantially similar to that of VPSIC, fails in the precise account of structural and magnetic
properties. In fact, our study revealed that subtle differences in electronic properties can result in visible errors in 
some observable quantities: specifically a slightly narrower p-d valence bandwidth and an excessive t$_{2g}$ localization
towards the lower end of the valence band manifold can result in a 2-3\% overestimation of the lattice constant and in a
unreliable value of the exchange-interaction parameter.

%%%%%%%%%%%%%%%%%%%%%%%%%%%%%%%%%%%%%%%%%%%%%%%%%%%%%%%%%%%%%%%%%%%%%%%%%%%%%%%%%%%%%%%%%%%%%%%%%
%%%%%%%%%%%%%%%%%%%%%%%%%%%%%%%%%%%%%%%%%%%%%%%%%%%%%%%%%%%%%%%%%%%%%%%%%%%%%%%%%%%%%%%%%%%%%%%%%
%%%%%%%%%%%%%%%%%%%%%%%%%%%%%%%%%%%%%%%%%%%%%%%%%%%%%%%%%%%%%%%%%%%%%%%%%%%%%%%%%%%%%%%%%%%%%%%%%
%%%%%%%%%%%%%%%%%%%%%%%%%%%%%%%%%%%%%%%%%%%%%%%%%%%%%%%%%%%%%%%%%%%%%%%%%%%%%%%%%%%%%%%%%%%%%%%%%

%===================================================================================================================
\section{Results: molecules}
\label{res_mol}
%=========================================================================================================================                                                                                                                  

\begin{table*}
\caption{\label{BLtab} Calculated bond-lengths of selected bonds from several molecules compared to experimental values. 
Both VPSIC and LSDA bond-lengths are shown. $\delta_{BL}$ in column 6 (7) denotes the absolute percentage difference between 
the calculated VPSIC (LSDA) and experimental bond-length}
\begin{ruledtabular}
\begin{tabular}{lcccccc}
Molecule&Bond&\multicolumn{3}{c}{Bond-length (\AA)}&\multicolumn{2}{c}{$\delta_{BL}~(\%)$}\\
\cline{3-5}\cline{6-7}\\
& &VPSIC&Experiment&LSDA&VPSIC&LSDA\\
\hline
BCl$_3$&B-Cl& 1.748 &  1.742 &  1.754 &  0.349 & 0.706   \\
CH$_4$&C-H&1.081&1.087&1.121&0.567&3.090\\
C$_2$H$_6$(Ethane)&C-C&1.478&1.536&1.524&3.751&0.776\\
C$_4$H$_8$(Cyclobutane)&C-C&1.503&1.555&1.548&3.315&0.419\\
C$_3$H$_6$(Cyclopropane)&C-C&1.471&1.501&1.519&1.986&1.168\\
O$_3$&O-O&1.224&1.278&1.273&4.200&0.364\\
NaCl&Na-Cl&2.317&2.361&2.336&1.883&1.039\\
SiH$_4$&Si-H&1.431&1.480&1.518&3.344&2.594\\
SiCl$_4$&Si-Cl&2.020&2.019&2.054&0.056&1.747\\
PH$_3$&P-H&1.407&1.421&1.460&0.997&2.736\\
PF$_3$&P-F&1.576&1.561&1.644&0.957&5.349\\
SH$_2$&S-H&1.341&1.328&1.382&0.984&4.030\\
CuF&Cu-F&1.763&1.745&1.735&1.009&0.548\\
ZnH&Zn-H&1.527&1.595&1.629&4.260&2.130\\
C$_2$H$_4$(Ethylene)&C=C&1.309&1.339&1.351&2.237&0.903\\
CO&C=O&1.124&1.128&1.153&0.384&2.217\\
CO$_2$&C=O&1.142&1.162&1.185&1.706&1.959\\
O$_2$&O=O&1.188&1.210&1.227&1.793&1.388\\
N$_2$&N$\equiv$N&1.086&1.098&1.119&1.077&1.924\\
C$_2$H$_2$(Acetylene)&C$\equiv$C&1.196&1.203&1.234&0.566&2.605\\
C$_6$H$_6$(Benzene)&C:C& 1.371 &  1.397 &  1.411 &  1.871 & 0.985   \\
\end{tabular}
\end{ruledtabular}
\end{table*}
%==============================================================================================

While the PSIC formulation was originally conceived to work for extended systems (i.e. within periodic boundary 
conditions) it may be no less useful for finite systems as well. Indeed, if for isolated atoms the full PZ-SIC 
is easy and straightforward, for large molecules and clusters its application may become mothodologically cumbersome and
computationally extensive, and the PSIC can furnish a practical and reliable alternative. The implementation in local 
orbital basis set and pseudopotentials (ASIC), carried out in Ref.\onlinecite{pemmaraju} in the framework of the SIESTA 
code\cite{SIESTA}, is ideally suited to the aim since it can treat both extended and finite systems on the same foot.
(In principle even the plane-wave implementation, albeit much less efficiently, can be applied to finite systems by 
supercell approach.) In the last few years a series of works related to molecules have been carried out by 
ASIC\cite{works_asic} with tipically satisfying results (provided that the relaxation parameter $\alpha$ is kept 
fixed to unity). The VPSIC implemented in local orbital basis set (i.e. the variational generalization of the ASIC)
is expected to yield KS spectra largely similar, although  not identical, to those obtained with ASIC. Additionally,
the performance of the VPSIC energy functional (eqn.~\ref{vpsic_etot}) and the associated forces (eqn.~\ref{forces}) 
for equilibrium molecular geometries needs to be investigated. 

In this section we look at the VPSIC description of the spectral and geometric properties of several molecules selected 
mostly but not exclusively from the standard G2 set~\cite{G2set}. Calculations were carried out using a development version
of the SIESTA code~\cite{SIESTA} within which the VPSIC method was implemented. Some details regarding the implementation 
are given in the appendix. For all of the atomic species, standard norm-conserving pseudopotentials generated using 
the Troullier-Martins scheme~\cite{TM} are employed including core-corrections where necessary. Scalar relativistic 
pseudopotentials are used for the III period elements. A numerical double-zeta-polarized (DZP) basis set~\cite{SIESTA} is 
employed for all of the atomic species. An energy shift of 50 meV is used to set the cut-off radius for the pseudo atomic 
orbital basis functions. Geometry optimizations are performed using a conjugate gradients algoritm until all of the forces
are smaller than 0.01 eV/$\AA$. 
%The Ceperley-Alder~\cite{CA} parameteization of the LSDA is employed in all of the calculations. 
 
%%%%%%%%%%%%%%%%%%%%%%%%%%%%%%%%%%%%%%%%%%%%%%%%%%%%%%%%%%%%%%%%%%%%%%%%%%%%%%%%%%%%%%%%%%%%%%%%%%%%%%
%%%%%%%%%%%%%%%%%%%%%%%%%%%%%%%%%%%%%%%%%%%%%%%%%%%%%%%%%%%%%%%%%%%%%%%%%%%%%%%%%%%%%%%%%%%%%%%%%%%%%%%%
\subsection{Equilibrium bond lengths}
%%%%%%%%%%%%%%%%%%%%%%%%%%%%%%%%%%%%%%%%%%%%%%%%%%%%%%%%%%%%%%%%%%%%%%%%%%%%%%%%%%%%%%%%%%%%%%%%%%%%%%%
%%%%%%%%%%%%%%%%%%%%%%%%%%%%%%%%%%%%%%%%%%%%%%%%%%%%%%%%%%%%%%%%%%%%%%%%%%%%%%%%%%%%%%%%%%%%%%%%%%%%%%%%

Table~\ref{BLtab} shows the equilibrium bond-lengths obtained within VPSIC of selected bonds in several gas phase 
molecules. The representative set chosen includes molecules mainly built from I, II and III period elements as well as 
non-magnetic transition metals and furthermore includes several species hosting different types of chemical bonds. 
Also presented for comparison are the corresponding LSDA and experimental bond-lengths. We find that the calculated VPSIC 
bond-length is generally shorter than the corresponding LSDA estimate. From column 5 in table~\ref{BLtab}, we see that the
LSDA bond-lengths are typically a few percent longer than those from experiment. On the other hand (see column 4 in 
table~\ref{BLtab}) VPSIC bond-lengths are seen to be a few percent shorter relative to experiment. In columns 6 and 7 we 
show the absolute percentage error $\delta_{BL}^i(X)=\frac{|L_i(X)-L_i^{expt}| \times 100}{L_i^{expt}}$ in the calculated 
bond-lenghts $L_i(X)$ relative to experiment $L_i^{expt}$ for each molecular species $i$ and $X\in\{$VPSIC,LSDA$\}$.
The estimated mean absolute percentage error over the test set $\Delta_{BL}$(X)=$ \frac{\sum_{i=1}^N \delta_{BL}^i(X)}{N}$
comes out to be 1.84$\%$ in LSDA and 1.77$\%$ in VPSIC. We also note further that within the test set, the maximum 
percentage error observed within VPSIC is $\sim4\%$ for the case of ZnH and that for the majority of molecules the error 
is typically under 2.0$\%$              

%%%%%%%%%%%%%%%%%%%%%%%%%%%%%%%%%%%%%%%%%%%%%%%%%%%%%%%%%%%%%%%%%%%%%%%%%%%%%%%%%%%%%%%%%%%%%%%%%%%%%%%%%
%%%%%%%%%%%%%%%%%%%%%%%%%%%%%%%%%%%%%%%%%%%%%%%%%%%%%%%%%%%%%%%%%%%%%%%%%%%%%%%%%%%%%%%%%%%%%%%%%%%%%%%%%%%
\subsection{Ionization potentials}
%%%%%%%%%%%%%%%%%%%%%%%%%%%%%%%%%%%%%%%%%%%%%%%%%%%%%%%%%%%%%%%%%%%%%%%%%%%%%%%%%%%%%%%%%%%%%%%%%%%%%%%%%%
%%%%%%%%%%%%%%%%%%%%%%%%%%%%%%%%%%%%%%%%%%%%%%%%%%%%%%%%%%%%%%%%%%%%%%%%%%%%%%%%%%%%%%%%%%%%%%%%%%%%%%%%%%

As in the case of the ASIC method, the primary advatage over LSDA afforded by VPSIC is expected to lie 
in the systematic improvement of KS eigenvalue spectra. The method is thus particularly relevant for DFT 
based electron transport schemes wherein an accurate description of the KS spectra \cite{cormac,smeagol} 
is often important. In exact KS DFT only the highest occupied orbital eigenvalue $(\epsilon^\mathrm{HOMO})$ 
has a rigorous physical interpretation and corresponds to the negative of the first ionization potential \cite{janak,PPLB}.
In general, for a $N$ electron system, the following equations hold in exact KS-DFT

\begin{equation}\label{discharge}
\epsilon^\mathrm{HOMO}(M) = -I_\mathrm{N}~~~~~~~\mathrm{for}~~( N-1 < M \leq N )
\end{equation}
\begin{equation}\label{charge}
\epsilon^\mathrm{HOMO}(M) = -A_\mathrm{N}~~~~~~~\mathrm{for}~~( N < M \leq N+1 )
\end{equation}
where $-I_\mathrm{N}$ and $-A_\mathrm{N}$ are the ionization potential (IP) and the electron affinity (EA) respectively. 
However, semilocal approximate functionals such as LSDA/GGA are known to perform poorly with regards to satisfying 
equations~\ref{discharge} and~\ref{charge} especially for molecular systems. In the following, we assess the mapping 
between electron removal or addition energies and the KS spectrum obtained from VPSIC also showing the corresponding
LSDA results for comparison. Furthermore, for both LSDA and VPSIC, the molecular geometries used to estimate the IP and 
EA are the corresponding equilibrium geometries in the neutral configuration. 

\begin{table}
\caption{\label{IPtab}Experimental Ionization potential (IP) compared to calculated negative HOMO eigenvalues for neutral 
molecules. Columns 2 and 3 present the results from LSDA and VPSIC respectively. The experimental
data are taken from reference \cite{expmolecules}.}
\begin{ruledtabular}
\begin{tabular}{lccc}
Molecule&\multicolumn{2}{c}{-$\epsilon^\mathrm{HOMO}$(eV)}&IP(eV)\\
\cline{2-3}\\ 
&LSDA&VPSIC&Experiment\\
& & &\\
\hline
BCl$_3$& 7.49 &  11.90 &  11.62 \\
CH$_4$&9.38&14.58&13.60\\
C$_2$H$_6$(Ethane)&8.11&12.81&12.10\\
C$_4$H$_8$(Cyclobutane)&7.41&11.80&10.70\\
C$_3$H$_6$(Cyclopropane)&7.23&11.69&10.60\\
O$_3$&7.66&13.57&12.73\\
NaCl&5.04&8.85&9.80\\
SiH$_4$&8.50&13.75&12.30\\
SiCl$_4$&7.97&12.46&12.06\\
PH$_3$&6.84&10.96&10.59\\
PF$_3$&8.32&12.96&12.30\\
SH$_2$&6.09&10.77&10.50\\
CuF&5.53&11.56&10.90\\
C$_2$H$_4$(Ethylene)&6.85&10.91&10.68\\
CO&8.80&13.91&14.01\\
CO$_2$&9.15&15.15&13.79\\
O$_2$&6.74&13.55&12.30\\
N$_2$&10.13&15.42&15.58\\
C$_2$H$_2$(Acetylene)&7.16&11.31&11.49\\
C$_6$H$_6$(Benzene)&6.63&  10.51&  9.23\\
\end{tabular}
\end{ruledtabular}
\end{table}

\begin{figure}[ht]
\includegraphics[width=0.45\textwidth, clip]{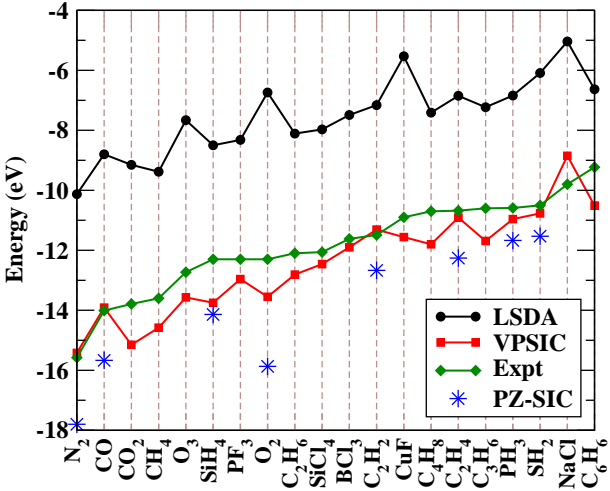}
\caption{\label{IPfig}Experimental negative ionization potential IP compared to the calculated HOMO eigenvalues 
for molecules. The experimental data are from reference \cite{expmolecules}, while the star symbol represents full
PZ-SIC calculations from reference \cite{scuseria2}.}
\end{figure}

In table \ref{IPtab} and figure \ref{IPfig} we compare the experimental IP for several molecules with the corresponding 
negative $\epsilon^\mathrm{HOMO}$ obtained using LSDA and VPSIC. It is clear that LSDA underestimates 
the removal energies significantly in all the cases. In contrast, the mapping between  the experimental IP and 
-$\epsilon^\mathrm{HOMO}$ from VPSIC is excellent. Indeed, the mean absolute deviation $\Delta_{IP}(X)$ ($X$ = LSDA, VPSIC)
from experiment, 
%
%\begin{equation*}
%\begin{eqnarray}
$\Delta_{IP}(X)=\frac{\sum_{i=1}^N |\epsilon^{\mathrm{HOMO},i}(X)+\mathrm{IP}^{i}_\mathrm{Expt}|}{N}$
%\end{eqnarray}
%\end{equation*}
%
is estimated to be 4.29~eV for LSDA and 0.72 eV for VPSIC~($N$ is the total number of molecules in table \ref{IPtab}). 
For comparison,  we have also included in figure \ref{IPfig} results obtained with a fully self-consistent PZ-SIC 
approach \cite{scuseria2}. Somewhat surprisingly the VPSIC approximation seems to produce better overall agreement with 
experiments than the full PZ-SIC scheme, which is seen to overcorrect the energy levels. This is a rather general 
feature of the PZ-SIC scheme and it has been suggested that some additional re-scaling procedure is 
needed~\cite{vydrov1,vydrov2}. 
%

%~~~~~~~~~~~~~~~~~~~~~~~~~~~~~~~~~~~~~~~~~~~~~~~~~~~~~~~~~~~~~~~~~~~~~~~~~~~~~~~~~~~~~~~~~~~~~~~~~~~~~
\begin{table*}[ht]
\caption{\label{EAtab}Calculated HOMO eigenvalues for singly negatively charged molecules compared to experimental 
negative electron affinities (-EA). Columns 6,7 and 8 present the LUMO eigenvalues for the corresponding neutral species. 
Experimental values are taken from ~\cite{scuseria2}}
\begin{ruledtabular}
\begin{tabular}{lccccc}
Molecule&\multicolumn{2}{c}{$\epsilon^\mathrm{HOMO}_\mathrm{N+1}$(eV)}&Exp. -EA (eV)
&\multicolumn{2}{c}{$\epsilon^\mathrm{LUMO}_\mathrm{N}$(eV)}\\
\cline{2-3}\cline{5-6}\\
&LSDA&  VPSIC &&LSDA& VPSIC\\
\hline
HC$\equiv$C$^-$	&1.79	&-2.60	&-2.97		&-6.54	&-7.59	\\
CH$_2$=CH$^-$	&4.13	&-0.19	&-0.67		&-3.49	&-5.07	\\
HC$\equiv$CO$^-$	&2.09	&-2.39	&-2.34		&-5.98	&-8.08	\\
CH$^-$	&3.80	&-0.82	&-1.24		&-4.81	&-7.15	\\
CH$_3^-$	&4.73	&-0.30	&-0.08		&-3.39	&-3.41	\\
CH$_3$O$^-$	&3.52	&-1.84	&-1.57		&-5.53	&-6.15	\\
CH$_3$S$^-$	&2.68	&-1.37	&-1.87		&0.3	&-5.73	\\
HC=O$^-$	&5.22	&1.05	&-0.31		&-3.36	&-6.86	\\
CN$^-$	&1.42	&-2.97	&-3.86		&-7.83	&-10.82	\\
CNO$^-$	&1.32	&-3.58	&-3.61		&-0.62	&-4.16	\\
%Cu2$^-$	&3.39	&1.86	&-0.84		&-2.32	&-3.46	\\
NH$_2^-$	&4.54	&-0.77	&-0.77		&-4.98	&-4.33	\\
NO$_2^-$	&4.64	&-0.93	&-2.27		&-5.02	&-9.47	\\
OF$^-$	&4.95	&-2.22	&-2.27		&-2.16	&-6.14	\\
OH$^-$	&4.46	&-1.52	&-1.83		&0.44	&-1.86	\\
PH$_2^-$	&2.94	&-0.98	&-1.27		&-4.55	&-4.99	\\
S$_2^-$	&2.83	&-0.01	&-1.67		&-4.5	&-6.67	\\
SH$^-$	&2.54	&-1.65	&-2.31		&-0.17	&-2.72	\\
SiH$_3^-$	&2.72	&-0.60	&-1.41		&-3.85	&-5.38	\\
%ZnH	&3.75	&0.98	&-0.94		&-2.9	&-4.82	\\
\end{tabular}
\end{ruledtabular}
\end{table*}
%~~~~~~~~~~~~~~~~~~~~~~~~~~~~~~~~~~~~~~~~~~~~~~~~~~~~~~~~~~~~~~~~~~~~~~~

%%%%%%%%%%%%%%%%%%%%%%%%%%%%%%%%%%%%%%%%%%%%%%%%%%%%%%%%%%%%%%%%%%%%%%
%%%%%%%%%%%%%%%%%%%%%%%%%%%%%%%%%%%%%%%%%%%%%%%%%%%%%%%%%%%%%%%%%%%%%%
\subsection{Electron affinities and the HOMO-LUMO gap}
%%%%%%%%%%%%%%%%%%%%%%%%%%%%%%%%%%%%%%%%%%%%%%%%%%%%%%%%%%%%%%%%%%%%%%%
%%%%%%%%%%%%%%%%%%%%%%%%%%%%%%%%%%%%%%%%%%%%%%%%%%%%%%%%%%%%%%%%%%%%%%%
%
\begin{figure}[ht]
\includegraphics[width=0.45\textwidth, clip]{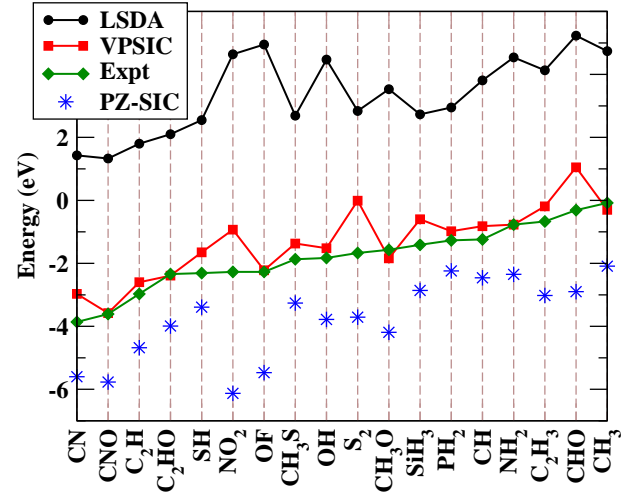}
\caption{\label{EAfig}Experimental negative electron affinities (-EA) compared to calculated HOMO eigenvalues 
of negative radicals.}
\end{figure}
In Hartree Fock theory, Koopmans' theorem~\cite{koop} implies that the lowest unoccupied molecular orbital (LUMO) 
energy $(\epsilon^\mathrm{LUMO})$, corresponds to the EA of the $N$ electron system when electronic
relaxation is neglected. No such physical interpretation exists for the Kohn-Sham $(\epsilon^\mathrm{LUMO})$ in DFT 
and so the EA is not directly accessible from the ground state spectrum of the $N$ electron system. 
However, as equation (\ref{charge}) indicates, the EA is in principle accessible from the ground state spectrum of 
the $N+1-f$ ($0 < f < 1$) electron system and furthermore, it must be relaxation free through non-integer occupation. 
Approximate functionals such as LSDA/GGA however perform rather poorly even in this regard as the $N+1$ electron state is 
often unbound with a positive eigenvalue. Therefore in practice, electron affinities are usually extracted either from more 
accurate total energy differences \cite{DSCF}, or by extrapolating them from LSDA calculations for the $N$ electron system
\cite{alessioaff}. The failure of approximate functionals in reproducing the spectra of anions has been traced in most 
part to the SI error and so SIC schemes are expected to perform better in this regard.

\begin{table}
\caption{\label{Gaptab}Calculated HOMO-LUMO gaps (E$_G$) of neutral molecules from LSDA and VPSIC. $\delta$E$_G$
in colum 4 represents the difference between corresponding VPSIC and LSDA gaps.}
\begin{ruledtabular}
\begin{tabular}{lccc}
Molecule&\multicolumn{2}{c}{E$_G$}&\\
\cline{2-3}\\ 
&LSDA&VPSIC&$\delta$E$_G$\\
%&    & & &\\
\hline
BCl$_3$	&4.84	&6.76	&1.92\\
C$_6$H$_6$(Benzene)	&5.33	&6.17	&0.84\\
C$_2$H$_2$(Acetylene)	&6.61	&8.27	&1.66\\
C$_2$H$_4$(Ethylene)	&5.58	&6.99	&1.41\\
C$_2$H$_6$(Ethane)	&9.03	&11.64	&2.61\\
%CH3-CH	&0.71	&4.31	&3.6\\
CH$_4$	&10.63	&13.79	&3.16\\
CO	&6.62	&9.41	&2.79\\
CO$_2$	&8.33	&11.03	&2.7\\
CuF	&1.5	&6.93	&5.43\\
C$_4$H$_8$(Cyclobutane)	&8.13	&10.29	&2.16\\
C$_3$H$_6$(Cyclopropane)	&8.15	&10.38	&2.23\\
N$_2$	&7.97	&10.27	&2.3\\
NaCl	&2.9	&6.75	&3.85\\
O$_2$	&2.16	&5.55	&3.39\\
O$_3$	&1.67	&3.23	&1.56\\
PF$_3$	&6.26	&9.11	&2.85\\
PH$_3$	&6.63	&8.47	&1.84\\
SH$_2$	&5.62	&8.0	&2.38\\
SiCl$_4$	&5.89	&7.79	&1.9\\
SiH$_4$	&8.64	&12.01	&3.37\\
TiO$_2$	&1.51	&4.01	&2.5\\
\end{tabular}
\end{ruledtabular}
\end{table}

In table \ref{EAtab} we compare HOMO energies (denoted as $\epsilon^\mathrm{HOMO}_{N+1}$) of several singly 
negatively charged molecules with the experimental electron affinities of the corresponding neutral species. 
We also report the LUMO energies for the molecules, most of which are radicals, in their neutral ground state 
(denoted as $\epsilon^\mathrm{LUMO}_\mathrm{N}$). Relaxed geometries of the neutral molecule are used for both 
the neutral and charged cases. We find that various $-\epsilon^\mathrm{HOMO}_\mathrm{N+1}$ obtained from VPSIC 
once again map quite well onto the corresponding experimental electron affinities in contrast to LSDA which yields 
unbounded states with positive $\epsilon^\mathrm{HOMO}_{N+1}$ for all the systems considered. Over the set of 
molecules in table~\ref{EAtab}, the mean absolute error with respect to experiment for the electron affinities 
%
%\begin{equation*}
\begin{eqnarray}
\Delta_{EA}(X)=\frac{\sum_{i=1}^N |\epsilon^\mathrm{HOMO,i}_{N+1}(X)+\mathrm{EA}^{i}_\mathrm{Expt}|}{N}
\end{eqnarray}
%\end{equation*}
%
($X$ = LSDA, VPSIC), stands at 4.67~eV  and 0.54~eV for LSDA and VPSIC respectively. 
%Notice that $\epsilon^\mathrm{HOMO}_\mathrm{N+1}$ from LSDA is positive in most cases as the states are unbound. 
In figure \ref{EAfig} we present our data together with $\epsilon_\mathrm{N+1}^\mathrm{HOMO}$ as calculated 
using the PZ-SIC \cite{scuseria2}. For EAs as well, we see that PZ-SIC seems to systematically
overcorrect the LSDA shortfall.

We now discuss briefly the HOMO-LUMO gap in VPSIC. As is apparent from columns 5 and 6 in table~\ref{EAtab}, the LUMO 
eigenvalues of the neutral species differ substantially from the corresponding negative electron affinities both within 
LSDA and VPSIC. In general DFT LUMO states are expected to be lower than -EA by an amount equal to the derivative 
discontinuity $\Delta_\mathrm{xc}$ defined as
\begin{equation}
\Delta_\mathrm{xc} = \lim_{f\rightarrow0^+}\epsilon^\mathrm{HOMO}_{N+f}-\epsilon^\mathrm{LUMO}_N\:,
\end{equation}
i.e. $\Delta_\mathrm{xc}$ is the discontinuity in the eigenvalue of the LUMO state at $N$. 
Thus the HOMO-LUMO gap is usually underestimated with respect to the true quasiparticle gap 
$E_g=I_\mathrm{N}-A_\mathrm{N}$. In table~\ref{Gaptab} we compare the HOMO-LUMO gaps from 
LSDA and VPSIC calculated in the neutral configuration for the molecular test set of 
table~\ref{IPtab}. We see that although the VPSIC HOMO eigenvalues are generally significantly 
lower than the corresponding LSDA ones (see table ~\ref{IPtab}), the HOMO-LUMO gaps differ 
by a smaller extent. This is because in contrast to other methods such as 
LSDA+U~\cite{AnisimovLongPaper} wherein the occupied levels are pushed lower and the 
unoccupied ones are pushed higher relative to the LSDA spectrum, within VPSIC the whole 
spectrum is corrected lower with the occupied and empty levels being shifted by different 
amounts depending upon their orbital character. For instance, the mean absolute difference 
between the LSDA and VPSIC HOMO-LUMO gaps for the test set in table~\ref{Gaptab} comes out 
to be $\sim$2.52 eV while the correction to the HOMO levels alone with respect to LSDA is 
around 4 eV. In general, VPSIC is expected to open the HOMO-LUMO gap substantially in systems 
where the occupied and un-occupied KS eigenstates have markedly different atomic orbital 
projections. Finally, it is worth noting that in contrast to explicity orbital dependent 
methods such as PZ-SIC, VPSIC does not exhibit the derivative discontinuity at integer 
occupations. Thus the eigenvalue of the highest occupied orbital relaxes continuously across 
fractional occupations although the range of eigenvalue relaxation is generally smaller than 
in LSDA. 

%%%%%%%%%%%%%%%%%%%%%%%%%%%%%%%%%%%%%%%%%%%%%%%%%%%%%%%%%%%%%%%%%%%%%%%%%%%%%%%%%%%%%%%%%%%%%%%%%%
%%%%%%%%%%%%%%%%%%%%%%%%%%%%%%%%%%%%%%%%%%%%%%%%%%%%%%%%%%%%%%%%%%%%%%%%%%%%%%%%%%%%%%%%%%%%%%%%%%%
%%%%%%%%%%%%%%%%%%%%%%%%%%%%%%%%%%%%%%%%%%%%%%%%%%%%%%%%%%%%%%%%%%%%%%%%%%%%%%%%%%%%%%%%%%%%%%%%%%
%%%%%%%%%%%%%%%%%%%%%%%%%%%%%%%    CONCLUSIONS   %%%%%%%%%%%%%%%%%%%%%%%%%%%%%%%%%%%%%%%%%%%%%%
%%%%%%%%%%%%%%%%%%%%%%%%%%%%%%%%%%%%%%%%%%%%%%%%%%%%%%%%%%%%%%%%%%%%%%%%%%%%%%%%%%%%%%%%%%%%%%
%%%%%%%%%%%%%%%%%%%%%%%%%%%%%%%%%%%%%%%%%%%%%%%%%%%%%%%%%%%%%%%%%%%%%%%%%%%%%%%%%%%%%%%%%%%%%%

\section{Conclusions} 
\label{concl}

In summary, we have introduced the first-principles VPSIC approach, a variational generalization of 
the method formerly known as PSIC/ASIC, based on the idea of removing the spurious self-interaction 
from the local-density functional energy. In VPSIC the self-interaction is removed in effective 
albeit approximate (i.e. orbitally-averaged) manner, which gives several advantages over the full SI 
removal (applied e.g. in the PZ-SIC approach or related methods for extended systems) such as the 
conservation of translational invariance (i.e. Bloch theorem) and the invariance of the energy under
unitary rotations of the occupied KS eigenfunction manifold. The VPSIC approach emerges as applicable 
to a vast series of systems (insulators and metals, magnetic and non-magnetic, extended or finite) 
with an overall satisfying accuracy, and furthermore not significally more demanding than LDA/GGA
from a computational viewpoint. 

We have implemented the method in two methodological frameworks: plane-waves basis set and ultrasoft
pseudopotentials, and local-orbital basis set plus norm-conserving pseudopotentials. The first was applied
here to extended systems: non-magnetic oxides, magnetic titanates, and magnetic manganites, the latter to a vast
range of molecules, testing structural and electronic properties. Overall, the performance of the VPSIC can be
synthesyzed as it follows: the predicted equilibrium structures are substantially in the same range of accuracy than
the LDA/LSDA, tipically reducing, on average, the bond lengths predicted by the latter. This corresponds to an
average underestimate of the experimental lattice constant by about 1\% for bulk systems, 2\% for molecule bond lenght.
On the other hand the electronic properties are highly improved with respect to LDA/GGA, with band gaps of bulks, and
ionization potentials and electron affinities for molecules tipically within 10\% from the respective experimental 
determinations. These preliminary results encourage us to pursue further explorations of the VPSIC approach for finite
and extended systems alike.

%%%%%%%%%%%%%%%%%%%%%%%%%%%%%%%%%%%%%%%%%%%%%%%%%%%%%%%%%%%%%%%%%%%%%%%%%%%%%%%%%%%%%%%%%%%%%%%
%%%%%%%%%%%%%%%%%%%%%%%%%%%%%%%%%%%%%%%%%%%%%%%%%%%%%%%%%%%%%%%%%%%%%%%%%%%%%%%%%%%%%%%%%%%%%%%
%%%%%%%%%%%%%%%%%%%%%%%%%%%%%%%%%%%%%%%%%%%%%%%%%%%%%%%%%%%%%%%%%%%%%%%%%%%%%%%%%%%%%%%%%%%%%%%%%
%%%%%%%%%%%%%%%%%%%%%%%%%%%%%%%%%%%%%%%%%%%%%%%%%%%%%%%%%%%%%%%%%%%%%%%%%%%%%%%%%%%%%%%%%%%%%%%%%
%%%%%%%%%%%%%%%%%%%%%%%%%%%%%%    APPENDICES  %%%%%%%%%%%%%%%%%%%%%%%%%%%%%%%%%%%%%%%%%%%%%%%%%%
%%%%%%%%%%%%%%%%%%%%%%%%%%%%%%%%%%%%%%%%%%%%%%%%%%%%%%%%%%%%%%%%%%%%%%%%%%%%%%%%%%%%%%%%%%%%%%%
%%%%%%%%%%%%%%%%%%%%%%%%%%%%%%%%%%%%%%%%%%%%%%%%%%%%%%%%%%%%%%%%%%%%%%%%%%%%%%%%%%%%%%%%%%%%%%%
%%%%%%%%%%%%%%%%%%%%%%%%%%%%%%%%%%%%%%%%%%%%%%%%%%%%%%%%%%%%%%%%%%%%%%%%%%%%%%%%%%%%%%%%%%%%%%

\section{Appendix I - Generalization of VPSIC formalism to USPP implementation}
\label{app_1}

For the study of large-size magnetic and strong-correlated systems, the ultrasoft pseudopotential 
(USPP) method\cite{uspp} associated to plane-wave basis set, represent a formidable tool which allows the use of cut-off
energies as low as 30-40 Ryd even for 'hard' transition metal ions such as Mn or Cu. The trade-off for this invaluable 
computational efficiency is a sensible complication of the VPSIC formulas presented in Section \ref{form}. 
In the following we furnish the VPSIC formulation adapted to the USPP formalism, which is our implementation of choice. 
For brevity however here we do not revisit the basic USPP formulation (for which we remand the readers to the original 
articles \onlinecite{uspp}), but only focus on the additional part specific for VPSIC.

In the USPP approach the atomic valence charge is partitioned in outer ultrasoft (US) and intra-core, augmented (AU) 
contributions, of whom only the first changes self-consistently with the surrounding chemical environment. Thus the charge
associated to the Bloch states $\psi_{n{\bf k}}^{\sigma}$ appearing in Section \ref{form} only represents the very 
smooth US part. The Bloch states obey the following generalized orthonormality conditions:

\begin{equation}
\langle \,\psi_{n{\bf k}}^{\sigma}\,|\, \hat{S}\,|\,\psi_{n'{\bf k}}^{\sigma}\,\rangle\,=\,\delta_{n,n'}   
\end{equation}

where the overal matrix $\hat{S}$ is given by:

\begin{equation}
\hat{S}\,=\,\hat{I} + 
\sum_{ab,\nu}\,|\beta_{a,\nu}\rangle q_{ab,\nu}\langle \beta_{b,\nu}|
\label{overl}
\end{equation}

Here $\beta_{a,\nu}({\bf r})$ and $q_{ab\nu}$ are atomic projector functions and augmented charge, 
respectively, characteristics of the USPP formalism, and ($a$,$b$) label atomic quantum numbers ($l_a$,$m_a$) 
($q_{ab\mu}$ $\neq$ 0 only for $l_a$=$l_b$). Consistently, the total charge density is generalized as:

\begin{equation}
n({\bf r})\,=\,\sum_{n{\bf k}\sigma}\, \langle \,\psi_{n{\bf k}}^{\sigma}\,|\, \hat{S}({\bf r})\,|\,
\psi_{n{\bf k}}^{\sigma}\,\rangle   
\end{equation}

\begin{equation}
\hat{S}({\bf r})\,=\,|{\bf r}\rangle \langle {\bf r}|+ 
\sum_{ab,\nu}\,|\beta_{a,\nu}\rangle Q_{ab\nu}({\bf r})\langle \beta_{b,\nu}|
\end{equation}

where $Q_{ab\nu}({\bf r})$ are augmented atomic charge densities. Within this generalized framework, the VPSIC energy
functional described in Eq.\ref{vpsic_etot} only includes the ultrasoft (US-SIC) part. In order to recover the full VPSIC
energy functional, a further augmented contribution must be added:

\begin{eqnarray}
E^{VPSIC-AU}\,=\,-{1\over 2}\sum_{ab\sigma\nu}\,B^{\sigma}_{ab,\nu}\,P_{ba,\nu}^{\sigma}{\cal E}^{AU}_{ba,\nu},
\label{eau}
\end{eqnarray}

where $B_{ab,\nu}^{\sigma}$ is the matrix of Bloch state projections onto the beta-function basis:

\begin{eqnarray}  
B_{ab,\nu}^{\sigma}\,=\,\sum_{n{\bf k}}\, f_{n{\bf k}}^{\sigma}\, \langle\psi_{n{\bf k}}^{\sigma}|
{\beta}_{a\nu} \rangle \langle\,{\beta}_{b\nu}|\psi_{n{\bf k}}^{\sigma}\rangle
\label{epsi}
\end{eqnarray}

and ${\cal E}^{AU}$ are SI energy associated to the augmented atomic charges $Q_{ab,\nu}({\bf r})$:

\begin{eqnarray}  
{\cal E}^{AU}_{ab,\nu}\,=\,\int\!\! d{\bf r}\,Q_{ab,\nu}({\bf r})\, V_{HXC}[n_{a\nu}({\bf r}),0].
\label{qij}
\end{eqnarray}

%The VPSIC-AU functional brings a corresponding additional contribution to the KS Equations:
%\begin{eqnarray}
%{\partial E^{VPSIC-AU}\over \partial \psi^*_{nk\sigma}}\,=\,
%-{1\over 2}\sum_{ab\sigma}\,
%\{ |\beta_{a\nu}\,\rangle\, Q_{ab,\nu}\langle\,\beta_{b,\nu}|\psi_{n{\bf k}}^{\sigma}\rangle \,P_{ba\sigma}\\
%%\end{eqnarray}
%%\begin{eqnarray}
%+|\phi_{a,\nu}\rangle \langle \phi_{b,\nu} |\psi_{n{\bf k}}^{\sigma}\rangle 
%\,B_{ba,\nu}^{\sigma} Q_{ij}^{SI} \} 
%\label{ks-aug}
%\end{eqnarray}

(In radial simmetry $V_{HXC}[n_{a\nu},0]$=$V_{HXC}[n_{b\nu},0]$ for $m_a$ $\neq$ $m_b$). Notice that we use different 
indices for the USPP projector ($a$, $b$) and for the SI projector ($i$, $j$), since the two basis set are conceptually 
and practically different: the latter is built on a minimal set of atomic orbitals, and to be physically sound it should
be associated to bound atomic states; on the other hand in order to improve the USPP transferability it is customary to 
include in the ($a$,$b$) matrix more than one state per angular moment (tipically the bound atomic state plusd one unbound
state relative to some diagnostic energy reference). This difference introduces some ambiguities in Eq.\ref{qij} relative 
to the definition of $V_{Hxc}[n_{a\nu},0]$ and P$_{ab\nu}$. The ambiguity can be solved by associating the same atomic 
$V_{Hxc}$ (relative to the bound state) to all the beta projectors with same angular momentum $l_a$. Another possibility 
is rewriting the VPSIC-AU energy as it follows:

\begin{equation}
E^{VPSIC-AU}[\{\psi\}]\,=\,-{1\over 2}\sum_{ij\nu\sigma}\, P^{\sigma}_{ij\nu}\,P^{\sigma}_{ji\nu}\,\epsilon^{AU}_{i\nu}
\label{eau-simp}
\end{equation}

where:

\begin{equation}
\epsilon^{AU}_{i\nu}\,=\,\sum_{ab}\,\langle\phi_{i\nu}\,|\,{\beta}_{a\nu}\,\rangle\,
{\cal E}^{AU}_{ab,\nu}\, \langle\,{\beta}_{b\nu}\,|\,\phi_{i\nu}\,\rangle
\end{equation}

is just the augmented-only SI energy relative to the atomic state $i$ at full occupancy, and can be directly calculated in
the atom. The use of the simplified Eq.\ref{eau-simp} bypass altogether the explicit presence of the augmented 
charges, thus greatly simplifying the VPSIC AU energy functional calculation. Since our many test cases reveal that 
Eqs.\ref{eau} and \ref{eau-simp} give indeed very similar results, we decide to adopt the latter as standard choice.  
Eq.\ref{eau-simp} brings a  corresponding contribution to the VPSIC KS Equations:

\begin{equation}
{\partial E^{VPSIC-AU}\over \partial \psi^*_{nk\sigma}}\,=
\,-\sum_{ij\nu}\, P^{\sigma}_{ij\nu}{\partial P^{\sigma}_{ji\nu} \over \partial \psi^*_{nk\sigma}}
\epsilon^{AU}_{j\nu} 
\label{ks-vpsic0_au}
\end{equation}

%%%%%%%%%%%%%%%%%%%%%%%%%%%%%%%%%%%%%%%%%%%%%%%%%%%%%%%%%%%%%%%%%%%%%%%%%%%%%%
Finally, the orbital occupations defined in Eq.\ref{occ} must be also generalized as:

\begin{eqnarray}
P^{\sigma}_{ij\nu}=\,\sum_{n{\bf k}}\, f_{n{\bf k}}^{\sigma}\, \langle\psi_{n{\bf k}}^{\sigma}|\tilde{\phi}_{i\nu}
\rangle \> \langle \tilde{\phi}_{j\nu} |\psi_{n{\bf k}}^{\sigma}\rangle,
\label{occ_norm}
\end{eqnarray}

where the ultrasoft atomic orbitals $\phi_{i\nu}$ have been replaced by

\begin{eqnarray}
|\tilde{\phi}_{i\nu} \rangle \,=\,\sum_{i'\nu'}\,S_{i'\nu',i\nu}^{-1/2}|\hat{S}\phi_{i'\nu'}\rangle
\label{phi_norm}
\end{eqnarray}

with $\hat{S}$ given in Eq.\ref{overl}. Thus, we have:

\begin{equation}
|\hat{S}\phi_{i\nu}\rangle\,=\,|\phi_{i\nu}\rangle + 
\sum_{ab,\mu}\,|\beta_{a\mu}\rangle q_{ab,\mu}\langle \beta_{b\mu}|\phi_{i\nu} \rangle
\label{overl_1}
\end{equation}

and

\begin{equation}
S_{i\nu,i'\nu'}\,=\,\langle\,\phi_{i\nu}\,|\,\hat{S}\,|\,\phi_{i'\nu'}\rangle
\label{sij}
\end{equation}

is the overlap matrix in the atomic orbital basis set. By construction this is Hermitian and also positive-definite, so its square 
root matrix can always be defined in the following, unique way: since (let's drop the atomic index) $[\hat{S}^{-1/2},\hat{S}]$=0, 
$\hat{S}$ can be diagonalized in the ($i$,$j$) subspace, and from its eigenvalues ${\lambda}_k$ we have $S^{-1/2}_{kk'}$= 
$\delta_{kk'}$ ${\lambda}^{-1/2}_k$. The latter is finally rotated back to the original ($i$,$j$) basis set to obtain $S^{-1/2}_{ij}$. 

The replacement of simple atomic-site centered $\phi_{i\nu}$ orbitals with the $\tilde{\phi}_i$ orbitals given by Eq.\ref{phi_norm} (known 
as L\"owdin orthonormalization \cite{lowdin}) is required by the necessity to enforce, at any $\{{\bf R}\}$, the orthonormality conditions 
$\langle \tilde{\phi}_{i\nu}| \tilde{\phi}_{j\nu'}\rangle$=$\delta_{ij} \delta_{\nu\nu'}$, and in turn, the correct normalization of the orbital 
occupation matrix defined in Eq.\ref{occ_norm}: (upon diagonalization) $0\leq P^{\sigma}_{ii\nu} \leq 1$ and 
$\sum_{i\nu\sigma}P^{\sigma}_{ii\nu}=$N, with N the total number of electrons in the cell. These constraints 
are essential to the interpretation of $P^{\sigma}_{ii\nu}$ as physically sound orbital occupancies.    

On the other hand, the L\"owdin renormalization implies a remarkable complicacy in the forces formulation, since 
it is clear from Eqs.\ref{phi_norm} and \ref{sij} that $\tilde{\phi}_{i\nu}$ is not simply centered on a single 
atomic site, but includes contributions from the overlap with all other atomic orbitals $\phi_{j\mu}$ as well as 
beta functions $\beta_{a\mu}$ of the cell. Since the forces formulation in the case of L\"owdin-normalized orbitals
may be useful even in other methodological contexts (e.g. in the LDA+U method, whose Hamiltonian is also written 
in terms of orbital occupancies) we dedicate the next Section to describe it in full detail.

%%%%%%%%%%%%%%%%%%%%%%%%%%%%%%%%%%%%%%%%%%%%%%%%%%%%%%%%%%%%%%%%%%%%
%%%%%%%%%%%%%%%%%%%%%%%%%%%%%%%%%%%%%%%%%%%%%%%%%%%%%%%%%%%%%%%%%%%%%
\section{Appendix II - Forces formulation within pane-waves and USPP}
\label{app_2}
%%%%%%%%%%%%%%%%%%%%%%%%%%%%%%%%%%%%%%%%%%%%%%%%%%%%%%%%%%%%%%%%%%%%%%
%%%%%%%%%%%%%%%%%%%%%%%%%%%%%%%%%%%%%%%%%%%%%%%%%%%%%%%%%%%%%%%%%%%%%%

In case of USPP formalism the forces espression given in Eq.\ref{forces} (or Eq.\ref{asi_forces}
if we the simplified approach is considered) must be generalized in order to include the contribution 
generated by the additional AU energy in \ref{eau-simp}: 

\begin{eqnarray}
-{\partial E^{AU}[\{\psi\}]\over \partial {\bf R}_{\nu}}
\nonumber
\end{eqnarray}

\begin{eqnarray}
=\sum_{ij,nk\sigma}\,f_{n{\bf k}}^{\sigma}\, \left\{ P^{\sigma}_{ij\nu} \epsilon^{AU}_{j\nu}\,
\langle\psi_{n{\bf k}}^{\sigma}| { \partial \phi_{j,\nu}\over \partial {\bf R}_{\nu} } \rangle 
\langle\,\phi_{i,\nu}|\psi_{n{\bf k}}^{\sigma}\rangle +c.c \right\}
\label{au_forces}
\end{eqnarray}

In plane waves basis set the implementation of Eqs.\ref{forces}, \ref{asi_forces}, or \ref{au_forces} is 
rather straightforward except for one ingredient which requires attention: the atomic orbital derivative.

The simplest case is that of atomic orbitals which remain centered on the atomic positions (i.e. orbitals which simply
translate along with their reference atom displacement). In this case the force on a given atom ${\bf R}_{\nu}$ 
only depends on the change of the orbitals sited on $\nu$, and the orbital derivative is easily calculated as:

\begin{eqnarray}
{\partial \over \partial {\bf R}_{\nu}} \langle {\bf k}+{\bf G}\,|\,\phi_{i\nu}\rangle 
= {\partial \over \partial {\bf R}_{\nu}} \, e^{-i \left( {\bf k}+{\bf G}\right)\cdot{\bf R}_{\nu} }\, 
\langle {\bf k}+{\bf G}\,|\,\phi_{i0}\rangle 
\end{eqnarray}

\begin{eqnarray}
= -i \left( {\bf k}+{\bf G}\right) \, \langle {\bf k}+{\bf G}\,|\,\phi_{i\nu}\rangle
\label{orb_der}
\end{eqnarray}

where clearly $\phi_{i\nu}$ = $\phi_{i\nu}({\bf r}-{\bf R}_{\nu})$, and $\phi_{i0}$ = $\phi_i({\bf r})$. However, 
as discussed in the previous Session, orbitals $\phi_{i\nu}$ must be replaced by $\tilde{\phi}_{i\nu}$, and the forces
equation generalized accordingly:

\begin{eqnarray}
-{\partial E^{VPSIC}[\{\psi\}]\over \partial {\bf R}_{\nu}}\,=\,{\bf F}_{\nu}^{LSD}+
\nonumber
\end{eqnarray}
\begin{eqnarray}
+{1\over 2}\sum_{ij,nk\sigma}\,f_{n{\bf k}}^{\sigma}\, \left\{ \langle\psi_{n{\bf k}}^{\sigma}| 
{ \partial \gamma_{i,\nu}\over \partial {\bf R}_{\nu} } \rangle 
C_{ij} \langle\,\gamma_{j,\nu}|\psi_{n{\bf k}}^{\sigma}\rangle P^{\sigma}_{ji\nu}\,+ c.c. \right\}
\nonumber
\end{eqnarray}
\begin{eqnarray}
+{1\over 2}\sum_{ij\mu,nk\sigma}\,f_{n{\bf k}}^{\sigma}\, \left\{ {\cal E}^{SI}_{ij\sigma\mu}\,
\langle\psi_{n{\bf k}}^{\sigma}| { \partial \tilde{\phi}_{i,\mu}\over \partial {\bf R}_{\nu} } \rangle 
\langle\,\tilde{\phi}_{j,\mu}|\psi_{n{\bf k}}^{\sigma}\rangle +c.c \right\}
\nonumber
\end{eqnarray}
\begin{eqnarray}
+\sum_{ij\mu nk\sigma}\,f_{n{\bf k}}^{\sigma}\, \left\{ P^{\sigma}_{ij\mu} \epsilon^{AU}_{j\mu}\,
\langle\psi_{n{\bf k}}^{\sigma}| { \partial \tilde{\phi}_{j\mu}\over \partial {\bf R}_{\nu} } \rangle 
\langle\,\tilde{\phi}_{i\mu}|\psi_{n{\bf k}}^{\sigma}\rangle +c.c \right\}
\label{au_forces2}
\end{eqnarray}
 
where the first two terms account for the US contribution, and the third for the AU part. 
The presence of L\"owdin-normalized orbitals brings one more sum over atomic positions in the second and 
third term, since now the displacement of one single atom in ${\bf R}_{\nu}$ changes in principles
all the orbitals, not just those sited on ${\bf R}_{\nu}$. The L\"owdin orbital derivatives gives: 

\begin{eqnarray}
\langle {\psi}_{n\bf k}\,|\, {\partial \tilde{\phi}_{j\mu}\over \partial {\bf R}_{\nu} } \rangle \,=\,
\nonumber
\end{eqnarray}
\begin{eqnarray} 
=\,S^{-1/2}_{j'\mu',j\mu} \langle {\psi}_{n\bf k}\,|\, {\partial {\hat{S}\phi}_{j'\mu'}\over \partial {\bf R}_{\nu}}\rangle
+{\partial  S^{-1/2}_{j'\mu',j\mu} \over \partial {\bf R}_{\nu} } 
\langle {\psi}_{n\bf k}\,|\, {\hat{S}\phi}_{j'\mu'} \rangle
\label{lowdin_der}
\end{eqnarray}

(where sum over repeated indices is understood). Let's start considering the first term: 

\begin{eqnarray}
\langle {\psi}_{n\bf k}|{\partial \hat{S}{\phi}_{j'\mu'} \over \partial {\bf R}_{\nu} } \rangle \,=
\langle {\psi}_{n\bf k}|\hat{S}| {\partial {\phi}_{j'\mu'} \over \partial {\bf R}_{\nu} } \rangle \,+
\langle {\psi}_{n\bf k}|{\partial \hat{S}\over \partial {\bf R}_{\nu}}| {\phi}_{j'\mu'} \rangle \,=
\nonumber
\end{eqnarray}
\begin{eqnarray}
\left[ \langle {\psi}_{n\bf k}| {\partial {\phi}_{j'\mu'}\over \partial {\bf R}_{\nu} } \rangle + 
  \sum_{ab\nu'} \langle {\psi}_{n\bf k}|{\beta}_{a\nu'}\rangle q_{ab\nu'} \langle {\beta}_{b\nu'}| 
 {\partial {\phi}_{j'\mu'}\over \partial {\bf R}_{\nu} } \rangle \right] \delta_{\mu'\nu} +
\nonumber
\end{eqnarray}
\begin{eqnarray}    
  \sum_{ab\nu'} [ \langle {\psi}_{n\bf k}|{\partial {\beta}_{a\nu'} \over \partial {\bf R}_{\nu} } \rangle 
     q_{ab\nu'} \langle {\beta}_{b\nu'}|{\phi}_{j'\mu'} \rangle + \nonumber \\
\langle {\psi}_{n\bf k}|{\beta}_{a\nu'} \rangle
     q_{ab\nu'} \langle {\partial {\beta}_{b\nu'}\over \partial {\bf R}_{\nu} } |{\phi}_{j'\mu'} \rangle ] 
\delta_{\nu'\nu}
\label{der_1}
\end{eqnarray}

Here the first term in square brackets select the contribution to the derivative due to the atomic orbital $\phi_{j'\mu'}$
displacement, while the second term select the contribution due to the beta functions displacement. Despite the apparent 
intricacy Eq.\ref{der_1} is rather straightforward to calculate in plane waves, since all these derivatives are easily
obtained through Eq.\ref{orb_der}. 

More involved is the calculation of the second term in Eq.\ref{lowdin_der}, which includes an uncommon square root matrix derivative.
We can proceed as following (dropping the atomic indices for brevity): from $S^{-1/2}\,S^{-1/2}$=$S^{-1}$ we obtain:

\begin{eqnarray}
{\partial S^{-1}\over \partial {\bf R}} \,={\partial S^{-1/2}\over \partial {\bf R}} \,S^{-1/2}+S^{-1/2}\,{\partial S^{-1/2} \over \partial {\bf R}}
\label{der_2}
\end{eqnarray}

The left side of Eq.\ref{der_2} can be transformed using the following espression:

\begin{eqnarray}
{\partial S^{-1}\over \partial {\bf R}} \,=\,S^{-1}{\partial S \over \partial {\bf R}} \,S^{-1}
\label{der_3}  
\end{eqnarray}

where $S^{-1}$ can be easily obtained from $S$ (just like $S^{-1/2}$, as explained in the previous Section). Then we need to calculate the
overlap matrix derivative (reintroducing atomic indices for clarity):

\begin{eqnarray}
{\partial S_{i\mu,i'\mu'}\over \partial {\bf R}_{\nu} }\,=\,\langle\,{\partial \phi_{i\mu}\over \partial {\bf R}_{\nu}}\,|\,\hat{S}\,|\,
\phi_{i'\mu'}\rangle
+\langle\,\phi_{i\mu}\,|\,\hat{S}\,|\,{\partial \phi_{i'\mu'}\over \partial {\bf R}_{\nu} } \rangle
\nonumber
\end{eqnarray}
\begin{eqnarray}
+\langle\,\phi_{i\mu}\,|\,{\partial \hat{S}\over \partial {\bf R}_{\nu} }\,|\,\phi_{i'\mu'}\rangle \,
\label{der_sij}
\end{eqnarray}

whose esplicit espression is clearly similar to what already reported in Eq.\ref{der_1}, and we can give it as understood. So, the matrix at the left 
side of Eq.\ref{der_2} is determined. Now looking at the right side, we see that if $S^{-1/2}$ commute with its derivative, the latter
is easily extracted as:

\begin{eqnarray}
{\partial S^{-1/2}\over \partial {\bf R}} \,={1\over 2} {\partial S^{-1}\over \partial {\bf R}} \,S^{-1/2} 
\label{der_4}
\end{eqnarray}

However, they do not generally commute (except when the Hermitian matrix $S^{-1/2}$ is also real) and Eq.\ref{der_4} does not hold.
Thus, we need to solve Eq.\ref{der_2}, which is nothing but a Lyapunov matrix Equation of the form $B$=$XA$+$AX$, where the known
terms are $A$=$S^{-1/2}$ and $B$=$dS^{-1}/dR$, and $X$=$dS^{-1/2}/dR$. The general Lyapunov Equation can be solved exactly, but 
for the specific values of $A$ and $B$ we can apply the simple strategy proposed in Ref. \onlinecite{zhu}, that we repeat here for 
the commodity of the reader: we can rewrite $A$=$CVC^+$, where $V$ and $C$ can be easily determined as the diagonalized matrix and 
the basis change matrix; then we can multiply both members of Eq. \ref{der_2} by $C^+$ on the left, and $C$ on the right: 

\begin{eqnarray}
C^+ B C = C^+ A X C + C^+ X A C = V C^+ X C + C^+ X C V
\label{der_5}
\end{eqnarray}

introducing $R$=$C^+ B C$ and $Y$=$C^+ X C$, Eq.\ref{der_5} can be rewritten as $R$ = $V Y$ +$Y V$, which is now trivially solved, 
given the diagonal character of $V$:

\begin{eqnarray}
Y_{ij}\,=\,{R_{ij} \over V_{ii}+V_{jj}}
\label{der_6}
\end{eqnarray}

Once $Y$ is determined, the unkonwn $X$ can be finally obtained as $X=C\,Y\,C^+$.

%%%%%%%%%%%%%%%%%%%%%%%%%%%%%%%%%%%%%%%%%%%%%%%%%%%%%%%%%%%%%%%%%%%%%%%%%%%%
%%%%%%%%%%%%%%%%%%%%%%%%%%%%%%%%%%%%%%%%%%%%%%%%%%%%%%%%%%%%%%%%%%%%%%%%%%%%%

\section{Appendix III - VPSIC formalism within atomic orbitals basis set}
\label{app_3}

%%%%%%%%%%%%%%%%%%%%%%%%%%%%%%%%%%%%%%%%%%%%%%%%%%%%%%%%%%%%%%%%%%%%%%%%%%%%
%%%%%%%%%%%%%%%%%%%%%%%%%%%%%%%%%%%%%%%%%%%%%%%%%%%%%%%%%%%%%%%%%%%%%%%%%%%%

As the VPSIC correction is based on a projection of the occupied KS orbital manifold onto a localized subspace of atomic orbitals, 
the formalism naturally lends itself to implementation within a localized orbital basis set framework. In this appendix we provide some 
details of the current VPSIC implementation within the SIESTA~\cite{SIESTA} code. The first step in setting up the VPSIC algroithm
is to construct a minimal set of atomic orbitals $\{\phi_{i,\nu}\}$ and the associated projectors $\{\gamma_{i,\nu}\}$ which
will be used to calculate the occupation numbers~(eqn.~\ref{occ}) and effective SI energies~(eqn.~\ref{epsi-si}). Within SIESTA, 
the functions $\phi_{i,\nu}$ are numerical pseudo-atomic orbitals with a finite range 
constructed as solutions of the atomic Schrodinger equation with an additional confining potential at the cutoff radius 
$r_\mathrm{c}$~\cite{SIESTA}. The finite extent of the functions $\phi_{i,\nu}$ also ensures that the 
corresponding SIC projectors $\gamma_{i,\nu}$ (eqn.~\ref{gamma}) also vanish beyond the cutoff $r_\mathrm{c}$.
In the current implementation, the SIC potential $V_{HXC}[\rho_{\nu,l_i}(r);1]$ in equation~\ref{gamma} is obtained
from a full PZ-SIC-LSDA~\cite{pz} calculation for a free atom and imported into SIESTA via a pseudopotential. An appropriate choice 
for the cutoff radius $r_\mathrm{c}$ is then dictated by the requirement that the expectation value 
\begin{eqnarray}
\delta\varepsilon_{i,\nu}^{SIC}=\int d{\bf r}\phi_{l_i,m_i}({\bf r})\,V_{HXC}[\rho_{\nu,l_i}(r);1]\,\phi_{l_i,m_i}({\bf r})
\label{cii}
\end{eqnarray}
reproduces the PZ-SIC-LSDA correction of the corresponding orbital in the free atom to within a small tolerance. 
Simultaneously, the cutoff should be reasonably short so as not to change the connectivity of the matrix elements of the PAO Hamiltonian. 
Therefore, in practice, we set the cutoff radius for the projection orbitals $\phi_{i,\nu}$ on a given atom to be either equal to the largest among
the cutoff radii of the PAO basis set for that particular atom (typically the first $\zeta$ of the lowest angular momentum),
or, if shorter, the radius at which $\delta\varepsilon_{i,\nu}^{SIC}<$~0.1mRy. For typical cutoff radii (6 to 9 Bohr),  we find that the atomic 
SIC-LSDA eigenvalues are reproduced to within 1 to 5 mRy for the most extended shells and to within 0.1 mRy for more confined shells. 
Thus $\delta\varepsilon_{i,\nu}^{SIC}$ are rather well converged already for cutoff radii defined by PAO energies shifts \cite{SIESTA} of around 20mRy.

Using the orbitals  $\phi_{i,\nu}$ and the projectors $\{\gamma_{i,\nu}\}$, the occupation numbers $p^{\sigma}_{ij\nu}$ and the 
effective SI energies $\epsilon^{SI}_{ij\sigma\nu}$ for the extended system can be calculated. Different choices are possible
for the projection operators that yield the occupation numbers $p^{\sigma}_{ij\nu}$. In our implementation we use the so called
\textit{dual} projection operator given by
\begin{eqnarray}
\hat{P}^{\sigma}_{ij\nu} = \frac{1}{2} \{ |\widetilde{\phi_{i,\nu}}\rangle\langle\phi_{j,\nu}|+|\phi_{i,\nu}\rangle\langle\widetilde{\phi_{j,\nu}}| \}
\end{eqnarray}
where $|\widetilde{\phi_{i,\nu}}\rangle$ is the dual orbital of $|\phi_{i,\nu}\rangle$ and is given by
\begin{eqnarray}
|\widetilde{\phi_{i,\nu}}\rangle = \sum_{j,\mu} S^{-1}_{i\nu,j\mu}  |\phi_{j,\mu}\rangle
\end{eqnarray}
with $S^{-1}$ being the inverse of the overlap matrix over the non-orthogonal set $\{\phi_{i,\nu}\}$
\begin{eqnarray}
S_{i\nu,j\mu} =  \langle\phi_{i,\nu}|\phi_{j,\mu}\rangle
\end{eqnarray}
The dual orbitals satisfy the orthogonality relation
\begin{eqnarray}
 \langle\widetilde{\phi_{i,\nu}}|\phi_{j,\mu}\rangle = \delta_{i\nu,j\mu}
\end{eqnarray}
With this choice for the occupation number operator, the matrix elements of the VPSIC potential between two basis functions $\alpha$,~$\beta$
come out as
\begin{eqnarray}
V^{SIC}_{\alpha\beta\sigma} = \frac{1}{2} \sum_{ij\nu} \epsilon^{SI}_{ij\sigma\nu} \{ \frac{1}{2} [ \langle\alpha|\widetilde{\phi_{i,\nu}}\rangle\langle\phi_{j,\nu}|\beta\rangle 
\nonumber
\end{eqnarray}
\begin{eqnarray}
+ \langle\alpha|\phi_{i,\nu}\rangle\langle\widetilde{\phi_{j,\nu}}|\beta\rangle ] \} + p^{\sigma}_{ij\nu} \langle\alpha|\gamma_{i,\nu}\rangle C_{ij\nu} \langle\gamma_{j,\nu}|\beta\rangle
\end{eqnarray}
The expression for the VPSIC contribution to the atomic forces also involves only two center integrals and their derivatives. Setting $S_{\alpha,i\nu} = \langle\alpha|\phi_{i,\nu}\rangle$,
$\tilde{S}_{\alpha,i\nu} = \langle\alpha|\widetilde{\phi_{i,\nu}}\rangle$ and $G_{\alpha,i\nu}=\langle\alpha|\gamma_{i,\nu}\rangle$,
\begin{eqnarray}
\mathbf{F}^{SIC}_{\mu} = -\frac{\partial E^{SIC}[\{\psi\}]}{\partial \mathbf{R}_{\mu}} 
\nonumber
\end{eqnarray}
\begin{eqnarray}
=\frac{1}{2}\sum_{ij\nu\sigma}\epsilon^{SI}_{ij\nu\sigma} \frac{\partial p^{\sigma}_{ij\nu}}{\partial  \mathbf{R}_{\mu}} + 
p^{\sigma}_{ij\nu} \frac{\partial \epsilon^{SI}_{ij\nu\sigma}}{\partial  \mathbf{R}_{\mu}}
\end{eqnarray}
with
\begin{eqnarray}
\frac{\partial p^{\sigma}_{ij\nu}}{\partial  \mathbf{R}_{\mu}} = 
\frac{1}{2}\sum_{\alpha\beta} \frac{\partial\rho^{\sigma}_{\beta\alpha}}{\partial\mathbf{R}_{\mu}} [\tilde{S}_{\alpha,i\nu}S_{\beta,j\nu}+S_{\alpha,i\nu}\tilde{S}_{\beta,j\nu}]
\nonumber
\end{eqnarray}
\begin{eqnarray}
+\rho^{\sigma}_{\beta\alpha}[\frac{\partial\tilde{S}_{\alpha,i\nu}}{\partial\mathbf{R}_{\mu}}S_{\beta,j\nu}+\tilde{S}_{\alpha,i\nu}\frac{\partial S_{\beta,j\nu}}{\partial\mathbf{R}_{\mu}}
\nonumber
\end{eqnarray}
\begin{eqnarray}
+\frac{\partial S_{\alpha,i\nu}}{\partial\mathbf{R}_{\mu}}\tilde{S}_{\beta,j\nu}+S_{\alpha,i\nu}\frac{\partial \tilde{S}_{\beta,j\nu}}{\partial\mathbf{R}_{\mu}}] 
\end{eqnarray}
and 
\begin{eqnarray}
\frac{\partial \epsilon^{SI}_{ij\nu\sigma}}{\partial  \mathbf{R}_{\mu}} = \sum_{\alpha\beta} \frac{\partial\rho^{\sigma}_{\beta\alpha}}{\partial\mathbf{R}_{\mu}} G_{\alpha,i\nu} C_{ij\nu} G_{\beta,j\nu}
\nonumber
\end{eqnarray}
\begin{eqnarray}
+\rho^{\sigma}_{\beta\alpha}[\frac{\partial G_{\alpha,i\nu}}{\partial\mathbf{R}_{\mu}}C_{ij\nu}G_{\beta,j\nu}+G_{\alpha,i\nu}C_{ij\nu}\frac{\partial G_{\beta,j\nu}}{\partial\mathbf{R}_{\mu}}] 
\end{eqnarray}
where the sum is over the SIESTA basis functions $\alpha$,$\beta$ and $\rho^{\sigma}_{\beta\alpha}$ is the density matrix given by
\begin{eqnarray}
\rho^{\sigma}_{\beta\alpha} = \sum_{n\mathbf{k}} f^{\sigma}_{n\mathbf{k}} \langle\beta|\psi^{\sigma}_{n\mathbf{k}}\rangle\langle\psi^{\sigma}_{n\mathbf{k}}|\alpha\rangle
\end{eqnarray}
\\

\section*{Acknowledgments}

Work supported by the European Union FP7 project under grant Agreement N. 233553 (project �ATHENA�) and N. 228989 (project �OxIDeS�); 
by the Seed project "NEWDFESCM of the Italian Institute of Technology;  by PRIN 2008 project "2-DEG FOXI", of the Italian Ministery of University and 
Research (MIUR); by the Fondazione Banco di Sardegna under a 2010 grant. Part of this work was carried out by A. F., V. F., and D. P. at the 2010 AQUIFER program 
of ICMR-UCSB in L�Aquila.Calculations performed at CASPUR Rome, Cybersar Cagliari.

%%%%%%%%%%%%%%%%%%%%%%%  biblio


\begin{thebibliography}{99}
\bibitem{note_iga}
Notice that we use a different reference configuration: in Ref.\cite{iga} they assume a local cubic reference system
where $z$ change according to Ti, so to be always parallel to the longer Ti-O bond. See our structural analysis later on.
\end{thebibliography}
\end{document}